\documentclass[twocolumn,aps,showpacs]{revtex4}
\usepackage{graphicx}
\usepackage{epsfig}
\begin{document}

\setlength{\unitlength}{1mm}
\bibliographystyle{unsrt} 
\title {Atomic interferometer measurements of Berry's and Aharonov-Anandan's phases 
\\ for  isolated spins   $S  > \frac{1}{2}$  non-linearly  coupled to external fields}
 \author{ Marie-Anne Bouchiat}
\affiliation{Laboratoire Kastler-Brossel, CNRS, UPMC, \'Ecole Normale Sup\'erieure,   
24, rue Lhomond, 75005 Paris France,}
\author{ Claude Bouchiat}
\affiliation{Laboratoire de Physique Th\'eorique de l'\'Ecole Normale Sup\'erieure, CNRS, UPMC,  
24, rue Lhomond, 75005 Paris France.}

\date {December 21, 2010}
\date{\today}
 \newcommand \be {\begin{equation}}
\newcommand \ee {\end{equation}}
 \newcommand \bea {\begin{eqnarray}}
\newcommand \eea {\end{eqnarray}}
\newcommand \nn \nonumber
\def \(({\left(}
\def \)){\right)}
 \def \va{{\mathbf{a}}}
 \def \vb{{\mathbf{b}}}
\def \vs{{\mathbf{s}}}
 \def  \vS{{\mathbf{S}}}    
 \def \vI{{\mathbf{I}}}
 \def \vr{{\mathbf{r}}}
 \def \vF{{\mathbf{F}}}
 \def \vr{{\mathbf{r}}}
\def \vp{{\mathbf{p}}}
\def \vE{{\mathbf{E}}}
\def \vX{{\mathbf{X  }}}
\def \vB{{\mathbf{B}}}
\def \ve{{\mathbf{e}}}
\def \vk{{\mathbf{k}}}
\def \vJ{{\mathbf{J}}}
\def \vC{{\mathbf{C}}}
\def \vD{{\mathbf{D}}}
\def \vd{{\mathbf{d}}}
\def \vSig{{\mathbf{\Sigma}}}
\def \vomega{{\mathbf{\omega}}}
\def  \veps{{\mathbf{\epsilon}}}
\def  \En{{\mathrm{E}}}
\def\bra{\langle}
\def\ket{\rangle}
\def\wt{\widetilde}
 \begin{abstract}

 In a recent article we have studied the peculiar features of the Berry and
Aharonov-Anandan's  geometric phases for isolated spins $ S\geq 1.$
We have assumed that they are submitted to a dipole and quadrupole 
coupling  to external $ \vE$ and $\vB$ fields  with the mild restriction $ \vE\cdot \vB=0$. 
This implies discrete symmetries  leading to remarkable simplifications of the 
geometry and algebra involved. The aim of the present work is to describe
realistic  proposals, within the realm of Atomic Physics, for the verification 
of some of our  most significant theoretical  predictions. There are several  
challenges to be overcome. For alkali atoms, most commonly used in
atomic interferometers, the only  practical way to generate quadrupole coupling,  
with a strength comparable to the dipole one, is the ac Stark effect induced 
by a nearly resonant light beam. One has then, to face the instability of
the``dressed" atom hyperfine (hf) level, candidate for our isolated spin.
One deleterious effect is the apparition of an imaginary part in the 
quadrupole to dipole  coupling strength ratio, $ \lambda$. Fortunately 
we have found a simple way  to get rid of $ \Im(\lambda)$ by an 
appropriate detuning. We are left with  an unstable isolated spin.
This implies an upper bound to the quantum cycle duration $ T_c $.
In the case of the Berry's phase,  $ T_c $ has a lower  bound coming 
from the necessity of keeping the non-adiabatic corrections below a
predefined level. We have found a compromise in the case of the
$ F=2 , m=0$    $^{87}$Rb ground state hf level. This is our candidate
for the measurement of the somewhat  ``exotic" Berry's phase acquired
by the $ S=2 , m=0$ state at the end of  a quantum cycle involving  a rotation
of $\pi$ of the $ \vE$ field - in practice the linear polarization  of the dressing beam -
about the $\vB$ field direction.  We have found a way to implement 
in a Ramsey-type interferometric measurement the  procedures aiming 
at a control of the non-adiabatic corrections, as    
described in details in our previous theoretical article. 
A numerical simulation of our experimental  
proposal  shows that  a $0.1\%$ accurate determination of Berry's phase, free of
non-adiabatic corrections, can be achieved. 
Measurements could be considered also for  
 cold $^{52}$Cr chromium atoms with $S=3$, where values of $ \lambda  \simeq 1 $
 can be obtained with an instability smaller than in the $^{87}$Rb case, due to a more    
favourable spectroscopic structure.
The  $ F=1, m=1$ hf  level of the  $^{87}$Rb ground state offers the opportunity
to extend the measurement of  Aharonov-Anandan's phases  beyond the case 
$ S=\frac{1}{2}$. We construct, using  ``light shift'', the Hamiltonian $H_{\parallel }(t)$ generating a closed 
circuit in the density matrix space which  satisfies  at any  time the 
``parallel transport" condition, thus making the quantum cycle free from the adiabaticity condition.  
 We also consider the case of half-integer spins  ({\it e.g.} $^{201}$Hg, $^{135}$Ba and $^{137}$Ba), with their own specific features. 
 We show how the difference of Berry's phases for states  $ S=\frac{3}{2}$ and  $S=\frac{1}{2 }$, with
 $m=\frac{1}{2 }$, can be exploited to achieve an holonomic maximum entanglement of three Qbits.
  
\end{abstract}
 \maketitle
\section{Introduction}

  In a separate paper \cite{bou3}  we have presented a theoretical study of the
Berry's phases  generated by cyclic evolution of isolated spins of arbitrary large values. It was assumed
that the spins interact non-linearly with time-dependent external electromagnetic fields (possibly
effective ones) via the superposition of a dipole and a quadrupole coupling. We have made the
assumption that the two effective fields are orthogonal, a mild restriction but with many advantages. It
implies several discrete symmetries of the spin Hamiltonian which simplify considerably the geometry and the algebra.
Our purpose  here,  is to suggest atomic physics experiments to observe the original features of the Berry's phases that we predicted, but are still not revealed.
   
  In the present work we shall be mainly concerned by the adiabatic quantum cycles
 within a given time  interval.  They are 
  generated by a Hamiltonian depending on a set of parameters  assumed to be a system of coordinates   for a  differential manifold. In this   way,
  an adiabatic quantum cycle generates a mapping of a close circuit drawn 
  upon  the  parameters space onto a closed loop  upon the density   matrix space. The Berry's phases can then be viewed as the geometric phases associated with  this particular class of quantum cycles.
    
    The quantum mechanics postulates imply that Berry's phases can be  written as  Bohm-Aharonov loop integrals. The associated Abelian gauge field is  acting  within  the  space 
formed by   the  external  parameters  of  the Hamiltonian  governing the quantum adiabatic cycles. In the case of a non-linear Hamiltonian, involving both dipole and quadrupole couplings with the  $\vB$ and $\vE$ fields satisfying the condition $ \vE \cdot \vB=0$,  the parameter space becomes isomorphic to the  two-dimension (2D) complex  projective space, $\vC P_2$. Quite remarkably,  $\vC P_2$ can be identified with  a solution of the Einstein equations  in the 4D real Euclidian curved space \cite{ Egu, Gib}. This kind of solutions appears 
in  Quantum Theory of Radiation under the name of ``Gravitional Instantons"\cite{GibHawk}. 
This strongly contrasts with the magnetic dipole case where the parameter space is the familiar 2D sphere, and  the Berry's phase gauge field the vector  potential  of a magnetic monopole,  written in spherical coordinates. When both the dipole and  the quadrupole couplings are present,  the  Berry's  gauge field has a more complex  structure,
 even for  quantum cycles lying upon   $\vC P_2$ subspaces isomorphic to the 2D sphere \cite{bou2}.  The observation  of these  non-trivial geometry features, predicted by  Quantum Mechanics, deserves, in our opinion, a precise experimental investigation. We suggest interferometric measurements, involving  light-shifted 
  $^{87}$Rb  hyperfine sub-levels, in order to exhibit these somewhat exotic quantum effects.

  It is quite natural to ask:  how can one get  from experiment 
      the geometric phase  associated with a quantum cycle of the density matrix  $ \rho $, since,
      by definition, $\rho $   is  phase   independent?
  The answer  is to be found within the Superposition Principle of Quantum   Mechanics: 
  any linear combination of two quantum states  $\vert \Psi_1 \ket $ and $\vert \Psi_2\ket $   relative to a given quantum system,
    $\vert \Psi_{1\,2}  \ket= c_1  \vert \Psi_1 \ket + c_2 \vert \Psi_2 \ket $ is  an accessible state for  the system.
    In the present atomic physics context, such a construction will be achieved via the interaction of the system with   specific   classical  radio-frequency fields, using so-called Ramsey pulses \cite{ram}. 
     The density
      matrix associated with $\vert \Psi_{1\,2}  \ket$ is 
  $ \rho_{1\,2} = \vert \Psi_{1\,2}  \ket \bra \Psi_{1\,2} \vert =\vert c_1  \vert ^2  \rho_1 +\vert c_2  \vert ^2  \rho_2
  + \Delta \rho_{1\,2} $. The crossed contribution: 
  $\Delta \rho_{1\,2}  =  c_1 \, c_2^{*} \vert \Psi_1  \ket\bra \Psi_2 \vert +h.c. $ contains all 
  the information needed to obtain the  difference of the geometric phases acquired by the states 
  $\vert \Psi_1 \ket $ and $\vert \Psi_2\ket $  during an adiabatic quantum cycle.
  We will discuss experimental schemes, where the  geometric phase 
  acquired by one state of the superposition will  be  known {\it  a priori} to be zero, and one measures then directly   the  phase acquired by the second state {\it  modulo }  $ 2 \pi $. 
  
  It is not possible to summarize here all the work stimulated by the original contributions of 
 M. Berry, B. Simon, Y. Aharonov, D. Bohm and J. Anandan \cite{Berry, simon,Bohm,AAphas}. We refer the reader to review 
papers \cite{wil, Anand}, pedagogical presentations \cite{SciAm, Holst,Pines}, several recent spin-off
in quantum computing \cite{Sjo,jones} and the possible impact on precision measurements
\cite{Comm,Pend,DeK,corn}.   But we have found in the literature only few papers 
 dealing with the Berrry's phase  for a spin submitted to a time-varying quadratic 
interaction \cite{tycko,pines1,wil}.   
The authors deal with the nuclear quadrupole resonance (NQR) spectra 
in a magnetic resonance experiment involving a rotating sample. However,
 they assume the absence of any magnetic interaction and this leads to  level-degeneracy. 
 Thus, the problem is generalized to the adiabatic transport of degenerate states. In such a situation, the
geometric  phase is replaced by a unitary matrix given by the Wilson loop integral of a $ SU(n)$ 
non-abelian gauge potential  where $n$ is the dimension of the eigenspace associated with
 a given degenerate quantum level \cite{wilzee, zee}. 
 This makes an important difference with respect to
 the conditions considered here, as well as in our theoretical work where level degeneracy was  
required to be absent. 
  
Throughout this paper we shall deal with a set of quadratic spin Hamiltonians
  \be
H(\vB(t),\vE(t))=\gamma_S \; \vS \cdot  \vB(t) + \gamma_Q\;  (\vS 
\cdot \vE(t))^2  \, , 
\label{themodel}
\ee
 containing a Zeeman shift produced by a magnetic field $\vB(t)$ (real or effective) and a quadratic Stark shift produced by an electric field $\vE(t)$ (mainly effective). The non-linear spin coupling is responsible for new physical features becoming apparent for $ S>1$. 
  
 A crucial step, in the  confrontation of the theoretical results with experiments,  was to realize that, rather than the standard dc-Stark effect,  
the ac-Stark shifts, induced by a nearly resonant linearly polarized laser beam, were  the  proper tools to generate an effective $\vE$ field. Samples of cold alkali atoms or trapped alkali-like ions appear then as good candidates for the observation of the Berry's phases induced by non-linear coupling. Our  spin system $ \vS $  is   identified  with the total   angular momentum $\vF $  acting upon  a given  hyperfine (hf)  sub-level  of an    alkali atom ground state.
 
 Our  method relies  upon   the second-order  ac Stark shifts involving optical   
frequencies nearly resonant for the transitions  $S_{1/2, F}\rightarrow P _{1/2, F^{\prime}}$, with an 
appropriate choice of the detunings \cite{cohen1}. This  would  then allow to use the hyperfine ground state sub-levels of  rubidium and cesium isotopes 
to simulate isolated spin systems  with   $ S $ having integer values between one and four.
 In addition, the laser frequency can be  tuned for making one of the   hyperfine sub-level
insensitive to the laser field. It will then  provide an absolute phase reference for the second  hyperfine sub-level
 which is performing a quantum cycle.

However,  the method of  ac-Stark shift to construct the   quadratic coupling,
 presents one drawback.  Under
the effect of irradiation by the light field,  the hyperfine ground state sub-level  used to 
simulate our spin system will  acquire a finite decay rate. This puts an upper limit upon the time available for performing one quantum cycle and detecting the associated phase shift. There are atomic species for which this detrimental effect is much less severe. For instance the $^{201}$Hg mercury isotope, (and alkali-earth-like atoms with half-integer nuclear spin) have a
 spectroscopic structure more favourable for this purpose. It  appears that $^{52}$Cr
chromium atoms with a spin 3 of purely electronic origin could provide   suitable and especially
interesting candidates.       

The derivation of the Berry's phase requires the validity of the adiabatic approximation. This means that to
perform a valid measurement, one must exert a very tight control upon the non-adiabatic
corrections, which are governed by the Hamiltonian-parameter velocities.  This  problem is
 particularly crucial, here, because of the instability introduced by the ac Stark shift in our  spin system. It is addressed in great detail in our previous theoretical work \cite{bou3}.
 We have proposed explicit solutions for approaching the adiabaticity criterion while ramping up the $\vE$ field and applying the angular rotation speed of the periodic Euler angles. We have given definite procedures  for reducing, eventually suppressing  non-adiabatic corrections of various  origins.  
 
As two examples of the new features of the Berry's phases generated  by the quadratic Hamiltonian let us  consider two particular set  of adiabatic cycles within the time interval $0\leq t\leq T_c$.

i) The $\vB $ field is precessing around a fixed axis, while the $\vE$ field (orthogonal to $\vB$)  is 
lying within the rotating plane defined by $\vB $ and the rotation axis.

ii) The direction of the $\vB $ field is fixed and the orthogonal $\vE$ field is rotating around $\vB$ by an angle $\alpha (t)$ with the boundary condition $\alpha (T_c) -\alpha (0) = \pi \;\;modulo \;\pi$. 
 Performing an   ``adiabatic'' increase of the $\vE$ field
 starting from a null value, we have  obtained  in \cite{bou3} a
mathematical form of the Berry's phase relative to cycle i)  which  looks  superficially
 similar to that of a linear  Hamiltonian. 
 However, there is  one {\it significant difference} : the contribution
involving the cosine of the B field tilt angle is no longer proportional to the magnetic quantum number
$m$ whatever the spin value, but instead  to the spin polarization along the $\vB$ field. In addition, {\it a new contribution} to the Berry's phase is
generated at the end of the cycle ii). It is given by the spin polarization times the rotation angular
speed $\dot{\alpha} (t)$ integrated over the $0\leq t \leq T_c$ interval.
A stricking case is that of $m=0$  for a spin larger than 1, the usual Berry's phase generated by
the linear Hamiltonian  has a null value, while the one generated by the quadratic Hamiltonian of Eq.
(\ref{themodel}) is non-vanishing and increases with the magnitude of  $\vE$ up to a maximum
growing with the value of the spin (especially when it is an even integer).  
Our aim, here, is to define the precise experimental procedure required for the observation of such new  features of the Berry's phase, emerging  from our  previous work \cite{bou3}. 

 We have already underlined that  the $\vE, \vB$ field orthogonality condition endows our problem with important symmetry properties and that it also makes the parameter space isomorphic to $\vC P^2$, but in addition, for the special case of spin one, this condition implies the isomorphism of the density matrix with the  parameter spaces. 
 Thus, the Berry's phase generated by  the quadratic spin Hamiltonian  was found  to be  mathematically identical to the Aharonov-Anandan
(AA) phase using an  appropriate parametrization  of $ \vC P^2 $.
 However, the physical contents are in general  different, 
 since, in contrast to the Berry's phase, the AA phase 
 is not restricted to {\it adiabatic} quantum cycles.
 This case was previously discussed theoretically by C. Bouchiat and G.W. Gibbons \cite{bou1}
 and C. Bouchiat  \cite{bou2}.  Using cold $^{87}$Rb atoms in
the hf state $F=1$ experiencing a quadratic ac Stark shift, we propose here to perform an experimental
comparison between the AA and Berry's phase in different adiabatic and
non-adiabatic regimes, first,  to verify, within the
adiabatic approximation, the identity between  Berry's and AA phases for  appropriate
parametrization and second, to observe the difference in their behaviour when the adiabatic approximation
begins to fail. We show how a geometric phase equal to Berry's should be still obtained by performing a quantum cycle within the condition of ``parallel transport'', $ Tr ( \rho(t)  \, H_{\parallel }(t))=0$, {\it without any constraint upon the time derivatives of  the physical observables}.
We shall discuss the realization of measurements of this kind with $^{87}$Rb atoms in their hf state $F=1$, showing interest for the fast accomplishment of elementary operations in quantum processing.      
{\it For spins larger than one}, the Berry's phase relative to the Hamiltonian considered in this work  loses the  identity with  the   AA  phase, which involves now
 closed circuits drawn upon  larger projective  complex planes  $\vC P^{2 S}$.
 
 Finally, as an additional support to the present experimental program, we would like to point out that it may have  other significative physical spin-off.  A first simple  suggestion would be to stop midway the adiabatic cycle described in the present paper.  For an appropriate choice of the Hamiltonian parameters, one obtains  an example of  ``coherent spin-squeezed"  states \cite{Kit,Sor,bou4}. A second example relies  upon the non-trivial  dependence upon $ \vS^2$ of  Berry's phases of  the present  paper, in contrast with  the dipole case where   Berry's phase  is  proportional  to the  magnetic quantum number $m$ whatever $S$. This  property is  the basic ingredient used   to perform an ``holonomic entanglement" of   $N$ non-correlated  spins 1/2  (or Q bits) having a  fixed number of spins ``down". The corresponding vector state can be  written as a linear combination of the eigenstates  $ \Psi_{S,M}^{i}$ of $ \vS^2$ and   $ S_z$  where  $ \vS = \sum_{i=1}^N\vs_i $  is   the total spin operator.  Despite the fact that  a given eigenvalue of  $ \vS^2$  may appear several times  among them,  the states   $ \Psi_{S,M}^{i}$    can  always be chosen   to have different symmetry properties upon the permutations of  the $N$ spins,  insuring their orthogonality. If  we assume that in formula (1)  $ \vS$  stands for the total spin operator introduced above, the  corresponding Hamiltonian $ H_N(t)$ is
  clearly invariant upon all the permutations of the $ N$ spins. As a consequence, it acts upon the states $ \Psi_{S,M}^{i}$   as if they were isolated spin systems. At the end  of the Berry' cycle, they acquire a phase depending upon their S values. This implies an entanglement of the  initially  non-correlated $N$ spin states \cite{bou3}. In the two particular examples with  $ N=3, 4$ and  one spin down, {\it {i.e} } $M=S-1$,  the parameters of the cycle can be chosen in order to achieve a maximum entanglement.  (The simplest non-trivial  example,  $N=3$,  is discussed  in the Appendix).

Sec.II is a summary of our theoretical work 
\cite{bou3} which presents the remarkable properties of Berry's phases acquired by an 
arbitrary spin, non-linearly coupled, at the end of a closed adiabatic quantum cycle. 
We also remind different procedures 
 for keeping the non-adiabatic corrections below a predetermined level.
 In Sec.III, the expression of the quadratic spin Hamiltonian Eq. (\ref{themodel}) is derived in terms of
the experimental parameters for alkali atoms lighted up  with a laser beam close to resonance. 
 We present new possibilities  offered by cold $^{52}$Cr chromium atoms
 which are now available.  
Sec.IV describes a Ramsey type interferometry measurement
of Berry's phase, free of any non-adiabatic correction, acquired by the $F=2, m=0$ hf 
substate of $^{87}$Rb, at the end of a quantum cycle induced by an $\vE$ field rotating around a 
fixed $ \vB $ field.  Sec. V  underlines  the  peculiar  features of half-integer spin Berry's phases, also yet non-observed. They are  illustrated  by  the case of spins three-half, where two types of measurements look possible.  
One involves $^{201}$Hg atoms (or other alkali-earth-like odd isotopes), the second a nuclear quadrupole resonance experiment on a uniaxial crystal placed in a magnetic field. 
Sec.VI deals with the particular case
S=1. We give a method to achieve parallel transport on $^{87}$Rb $F=1$ atoms and perform an experimental comparison between Berry's and AA phases. Finally Sec.VII  is a summary  of our work and possible development. 

\section{Summary of Berry's phase theory 
for  arbitrary spins non-linearly coupled to external fields  } 
Throughout this paper, we make two important assumptions, first {\it the two spin couplings}, linear and quadratic, can be made {\it of comparable magnitudes}, second {\it the effective fields $\vB$ and $\vE$ are orthogonal}. 
\subsection{ Symmetry properties of  the non-linear spin Hamiltonian.  Physical implications}
In order to exhibit  the geometric structure of the  Hamiltonian $ H(\vB,\vE ) $, let us introduce the rotation $ R(t) $ mapping the coordinate   trihedron $ (\hat x, \hat y,\hat z)$ upon 
the trihedron defined by the directions of the $ \vB, \vE$ fields:
 $ R(t) = {\cal R}(\hat z,\varphi(t)){\cal R}(\hat y,\theta(t)){\cal R}(\hat z,\alpha(t))$,
where     ${\cal R}(\hat u ,\chi(t)) $ is standing  for the rotation of angle $ \chi(t)$ around the unit vector  $\hat u$. A very convenient  mathematical tool  is the unitary transformation:
  $ U(R(t)) = \exp -\frac{i}{\hbar} S_z \varphi(t)   \exp -\frac{i}{\hbar} S_y \theta(t) 
   \exp -\frac{i}{\hbar} S_z \alpha(t)  $, which  performs the rotation of the spin operator:
  $ U^{\dagger}(R(t)) \,\vS \,U(R(t)) = R(t)\cdot \vS $. 
  This rule   allows us to write: $ H(\vB, \vE)= \gamma_S B \hbar \; U(R(t)){\cal H}(\lambda )U^{\dagger}(R(t))$, with   ${\cal H}(\lambda ) = S_z \hbar^{-1} + \lambda \,S_x^2 \hbar^{-2}$. Taking aside the energy scale  $\gamma_S B\hbar$,   the  dimensionless  parameter $\lambda = \hbar \gamma_Q E^2/ (\gamma_S B)$, combined with the Euler angles $\theta, \varphi, \alpha$  provides  a  set  of  dimensionless  parameters for  $  H(\vB,\vE)$, which  could serve  as  coordinates for the $\vC P^{2}$ projective space.
  
  The eigenstates of ${\cal H}(\lambda )$, $\hat \psi (\lambda, m)$    are labeled with a ``magnetic'' number $m$ by requiring that their analytical continuation towards $\lambda=0$  coincide with  the  angular momentum eigenstates  $\vert S, m\ket$.
 Similarly, the associated eigenenergies  $ {\cal E}(\lambda, m) $ satisfy the boundary 
 condition: $  {\cal E}(0, m)=m$.
 
  Thanks to the constraint $\vE\cdot \vB=0$,  ${\cal H}(\lambda )$ has several discrete symmetries, leading to important physical consequences \cite{bou3}.
  \begin{itemize}
  \item {\it a.}
  The ``$m$'' parity $(-1)^{S-m}$, associated with a $ \pi$-rotation around $\hat z$,  is a good quantum number for ${\cal H}(\lambda )$. As a consequence, ${\cal H}(\lambda )$ cannot  mix angular momentum states $\vert S, m\ket$ with opposite ``$m$'' parities and it 
 can be expressed, within  the angular momentum basis, as  the  direct sum   of two matrices  acting respectively upon the states  with even and odd  $m$ parity: 
 $ {\cal H}(\lambda)=   {\cal H}_{even}(\lambda)  \oplus  {\cal H}_{odd}(\lambda).$ 
    \item {\it b.}
    ${\cal H}(\lambda)$ obeys  under the  rotation   ${ \cal R}(\hat x ,\pi) $   
    the transformation law:
      ${\cal H}(\lambda)  \rightarrow  - {\cal H}(-\lambda). $   Since 
      the rotation  ${ \cal R}(\hat x ,\pi) $
      flips the spin  component along the $z$ axis, one gets the symmetry relation:
    ${\cal E}(m,\lambda) = -{\cal E}(-m, - \lambda)$.  
    \item {\it c.}
    The invariance of ${\cal H}(\lambda)$ upon the rotation   ${ \cal R}(\hat z,\pi) $   
    implies that the quantum average of $ \vS $ relative to the  vector state 
     $\hat \psi (\lambda, m)$ lies along the $z$ axis:
    $ \langle  \hat \psi (\lambda, m) \vert \vS  \vert   \hat  \psi (\lambda, m)  \rangle=
    \hbar \, p(m,\lambda) \hat z $ where $p(m,\lambda)$ will be referred as the polarization of the spin state.  Remembering that the unit vector  $\hat z$  is taken  along the $\vB $ field, one gets for the following  quantum 
    spin average:
       $ \bra\,  \vS  \,\ket = \hbar \, p(m,\lambda) \,\vB/B $.
    \item {\it d.} 
    The last invariance is somewhat more subtle  than the three others  since 
 it   involves  the antiunitary  transformation associated with the product of the time reversal 
 by the  space reflection  with respect to the $x y$ plane. It implies that
 $ \hat \psi (\lambda, m) $ {\it can be represented by a real vector}. As we shall see this result has 
 non-trivial physical consequences. 
  \end{itemize} 
    \begin{figure*}
\centering\includegraphics[width=15 cm]{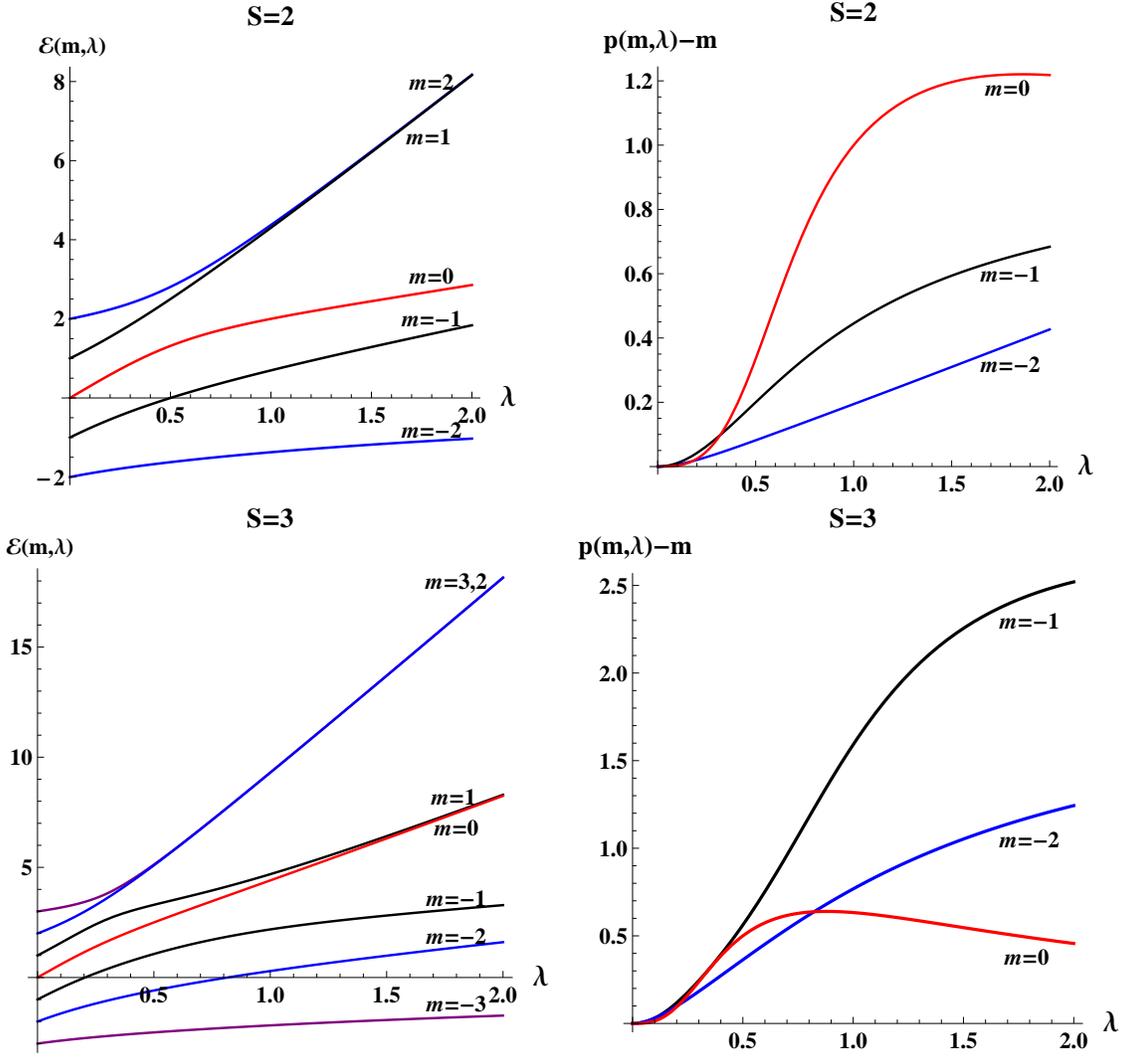}
\caption{ \small (Color online) Reduced  energies ${\cal E}(m,\lambda)$ (left-hand) and gauge field components $A_{\alpha}= p(m,\lambda) -m $ (right-hand)  versus $\lambda$ for $S=2$ and $ S=3$, for selected $m$-values relevant for the discussion presented in subsec. 2.C. 
Intersection of the energy curves with the vertical axis indicates the eigenvalues $m$ of $S_z$ for $\lambda=0$. For $\lambda >0$ remarkable effects appear for $S=2,\; m=0$ and for $ S=3, \; m=-1 $. In both cases this occurs in a range of Stark to Zeeman coupling ratios where the levels are well separated.} 
 \label{fig1}
\end{figure*}
\begin{figure*} 
  \centering\includegraphics[width=14.5 cm]{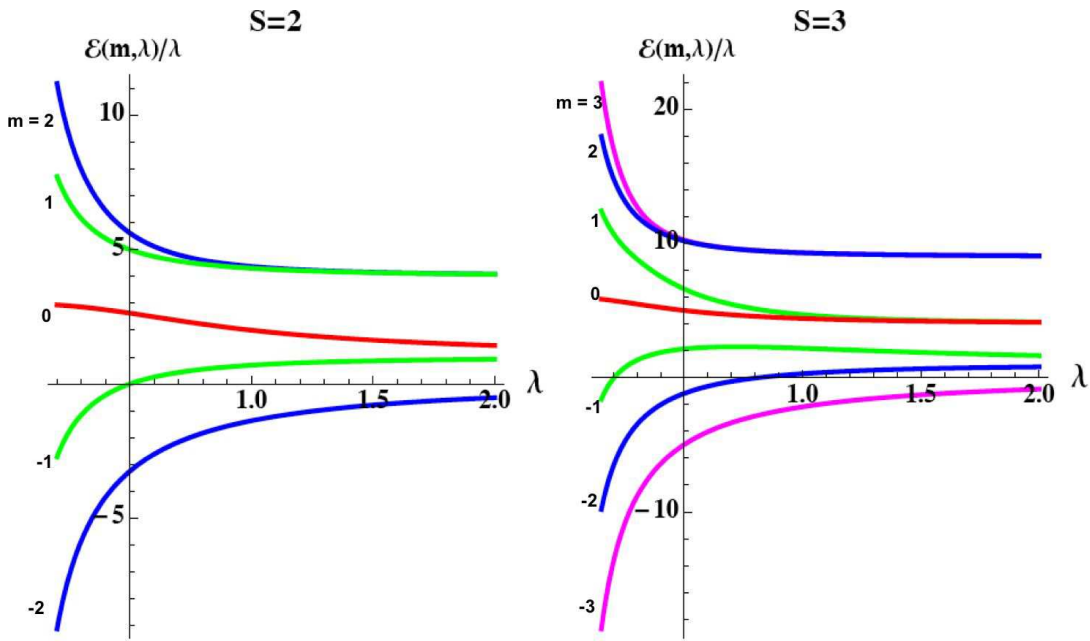}
  \caption{ \small (Color online) Plot of $ {\cal E}(m, \lambda)/\lambda $  within the interval $ 0.1 \leq \lambda \leq 2$, 
allowing us   to clarify the rearrangement of $ 2 S+1$ non-degenerate levels for $\lambda
\ll1$, into $ S$ degenerate doublets  with ${\cal E}(m, \lambda)/\lambda \simeq \mu^2$ in  
the  limit $ \lambda  \gg 1$,  $\mu$ being an integer such that $-S\leq \mu \leq S$.  Starting from the even-odd (or odd-even) pair:
 $ \{  {\cal E}(S, \lambda)/\lambda$,  ${\cal E}(S-1, \lambda)/\lambda \}$,  one sees
 clearly that the pair converges, {\it without  crossing},  towards the degenerate doublet with:
   $ {\cal E}(S, \lambda)/\lambda \simeq  {\cal E}(S-1, \lambda)/\lambda  \simeq   S^2 .$
 The next lower pair will  end as the degenerate doublet having the energy $ \simeq  \lambda \, 
(S-1)^2 $ and so on, until one reaches the  isolated level with  $ \mu= - S$.  It  has  no other 
possibility  than to converge to the non degenerate  level with  $ \mu^2 =0$. The fact that 
there is no level crossing  for finite values of $ \lambda $ follows from the very simple 
mathematical structure of  the Hamiltonian $ \hat{ \cal H} ( \lambda )$. Indeed,
   any modification of its symmetry properties,  is excluded  until one reaches 
   the limit $ \lambda \rightarrow  \infty $. }   
 \label{fig2}
\end{figure*}
  The eigenvalues of $\ H(\vB,\vE)$ together with  the associated  eigenvectors   $ \Psi (m,\lambda)$ can be  written as:
$E(m,B,E)= \gamma_SB \; {\cal E}(m,\lambda)\, ; \, \Psi(m,\lambda,t)= U(R) \hat \psi(m,\lambda).$
The  spin quantum average relative to $ \Psi (m,\lambda)$  is along the  $ \vB $ direction:
$\langle \vS \rangle = \hbar  \, p(m,\lambda) \vB/B$. Using the Hellmann-Feynmann theorem,  $ p(m,\lambda) $  is obtained by taking the partial  derivative of the energy  $ E(m,B,E)$   with respect to $\gamma_S B$:
$p(m,\lambda)= {\cal E}(m,\lambda) - \lambda \frac{\partial {\cal E}(m,\lambda)}{\partial \lambda}$.
Note that $ p(m,\lambda)$ obeys   under the reversal of $\lambda$  the
same  symmetry law as ${\cal E}(m,\lambda)$ derived   in $ \bullet \, b.$
 The curves representing  the variations of ${\cal E}(m,\lambda)$ and $p(m, \lambda)$ versus  $\lambda$ for $ S=2,3$ and 4 are displayed in  Fig. 1 of our previous paper  \cite{bou3}.
 \subsection{ A sketchy derivation of Berry's phases for  
  adiabatic cycles  governed  by  $\ H(\vB,\vE)$ }
 We have now all the necessary  ingredients  to sketch the derivation of the Berry's phase
 using  the instantaneous  eigenfunctions $ \Psi(m,\lambda,t) $ given in the
 previous subsection. Our starting point is the standard formula:
 \bea 
 \beta(m)&= & \int_0^T   dt \bra\Psi(m,\lambda,t)\vert  \, i\, \frac {\partial}{\partial t} 
\Psi(m,\lambda,t)\ket+\phi(m)   \nonumber \\
{\rm with} \;\;\phi(m) &=&\arg\((\Psi(m, \lambda(0),T)/\Psi(m,0,\lambda(0) \)),
\label{formphas}
 \eea  
 where we have used the fact that $ \lambda $ is a non periodic parameter:
  $\lambda(T)= \lambda(0)$.
  Inserting $\Psi(m,\lambda,t)= U(R) \hat \psi(m,\lambda)$,  the time 
  integral contribution  can be rewritten as: \\
  $ \gamma(m)= 
  \int_0^T  dt  \, i\,\hbar \bra \hat \psi(m,\lambda)\vert \((  \frac {\partial}{\partial t} 
   + U^{\dagger}(R) \frac {\partial}{\partial t} U(R)  \)) \hat\psi(m,\lambda)\ket.        $
Since $\hat \psi(m,\lambda)$ is a normalized real vector,   
the first term inside the parenthesis vanishes. In the second term, the operator  
$   \, i\,\hbar U^{\dagger}(R) \frac {\partial}{\partial t} U(R) $ is a standard
group theory object which can be expressed under the canonical form: 
$  \, i\,\hbar U^{\dagger}(R) \frac {\partial}{\partial t} U(R)= \vec{\omega}(t) \cdot \vS$, 
where  $\vec{\omega}(t)$ is a real vector. Its three components are linear functions of 
the time derivatives of the Euler angles. Their explicit expressions can be found 
in references  \cite{bou1,bou3}. Using $ \bullet\, c$, one sees immediately that only the z- component $ \omega_z = \cos(\theta) \dot{\varphi}+ \dot{\alpha}$
 does contribute: 
 $\gamma(m)=\int_0^T  dt \,  p(m,\lambda) \,(  \cos(\theta)\, \dot{\varphi}+ \dot{\alpha}).$
The calculation of $\phi(m)$ is performed in details in  references  \cite{bou3}. 
It is greatly simplified by the  particular features of the  expansion of  $\hat\psi(m,\lambda)$ over the angular  momentum eigenstates  resulting from properties $\bullet\, a.$ and $\bullet\, d$, which lead to :\\
$ \vert \hat \psi(m, \lambda(0))=\sum _{\vert m- 2n \vert \leq S }
  C_{m,n}(\lambda(0)) \vert S, m-2 n \ket.$
The sum runs upon states of the same m-parity and the coefficient
$C_{m,n}$ are real numbers. One can factorize out from  $ U(R(T)) U^{\dagger}(R(0))$
the operator  giving   the phase shift $\phi(m) $:
$ U_{\phi}=  \exp- \frac{i}{\hbar} S_z  \((\varphi(T) - \varphi(0) +\alpha(T)-\alpha(0)\))$ 
Remembering  the quantum cycle boundary conditions:
 $  \varphi(T) - \varphi(0)= 2 n_{\varphi}\,\pi\, , \alpha(T)-\alpha(0)= n_{\alpha}\,\pi  $,
 one sees that the effect of  $U_{\phi}$ upon   the terms of the  expansion of
$ \hat \psi(m, \lambda(0))$  is  just to multiply them by the same phase factor:
$ \exp- i \,m   \((\varphi(T) - \varphi(0) +\alpha(T)-\alpha(0)\))$. In this way one  
obtains  $\phi(m) =- \int_0^T  m (\dot \varphi (t) +\dot \alpha (t)) \, dt $. 
 
We arrive at our final expression for the  Berry's phase in terms
of a loop integral  along a closed circuit drawn upon the parameter space $\vC P^2 $:
\bea
\ \beta(S,m)&=& \oint_{{\cal C }} A_{\varphi}(S,m,\lambda)\, d\varphi +
A_{\alpha }(S,m,\lambda)  \, d\alpha, 
  \nonumber  \\
A_{\varphi}(S,m,\lambda) &=& p(S,m, \lambda) \cos \theta -m,   \nonumber \\ 
 A_{\alpha }(S,m,\lambda) & =&  p(S,m, \lambda)-m.   \label{BPphase}
\eea 
In the writing, we have stressed the two new  physical features 
introduced by the quadrupole spin  coupling: first,   
 the  Aharonov Bohm-like  integral now involves a two-component  Abelian gauge  
field $ (A_{\varphi}, A_{\alpha} )$  instead of a single one, second,  it now exhibits 
a strong dependence upon the value of $ \vS^2= \hbar^2 S (S+1) $, in contrast
with the dipole case where the Berry's phase  depends only upon  $m$. The latter 
effect reflects the fact  that the polarization $p(S,m, \lambda)$ has a simple 
  linear relation with ${\cal E}(m,\lambda)$.  These eigenenergies   are given by  the roots 
  of  two polynomials having respectively the degree $ S+1$ and $ S-1$ and   
  coefficients which are  $\lambda$  monomials. As a consequence, for $ S\geq 4$ the eigenenergies 
  are given by transcendental functions of  $\lambda$. 
 \subsection{ Physical implications of the S-dependence of Berry's phases
 generated by $ H(\vB,\vE) $}
 For integer spin values, remarkable effects  are predicted when  Berry's cycles  have as  initial state the  angular momentum eigenstate $ \vert S ,0 \ket $, corresponding to 
an  initially  vanishing polarization $ p(S,0,0)=0$. If  no $\vE$ field is applied
during the cycle, the  polarization keeps its null value all along the cycle and 
one recovers the well known result $ \beta(S,0,0)=0.$
 However, if  the  $\vE $ field  intensity is ramping up  to  a maximum before $\alpha$-rotation starts and returning 
 to a null value when it stops,   
 the inital state  is mixed    with states  $\vert S,m\ket $ with  $ m= \pm 2 n $ (n is an integer      
  $\leq S/2$).  For $S>1$ a finite polarization  
  $ p(S,0,\lambda)$ appears to third order in $\lambda$ given by
 $p(0,\lambda) = \frac{1}{8}\lambda^3S(S+2)(S^2-1)(1+{\cal O}(\lambda^3))$.
   Finite  values for  the gauge field  $ (A_{\varphi}, A_{\alpha} ) $ are then generated.
   
    In the following we concentrate upon the cycles  where
$ \alpha $ and $\lambda$ are the sole varying parameters. As a consequence, only the gauge field component $A_{\alpha }$    is relevant.
We have displayed in Figure 1 the variations  with $ \lambda$  of ${\cal E}(S,m,\lambda)$ for $ S=3$ and 2 with $ m= 0, \pm 1, \pm 2 $, together with  that of the gauge field 
 $ A_{\alpha } ( S,m, \lambda) $ for $m=0,-1,-2$. One sees clearly that
  the relevant energy levels are well separated.
   The magnitude of $ A_{\alpha }(S;m, \lambda)$ is remarkably large for  $S=2, m=0$,  in strong contrast with the case  of  a  linear spin  Hamiltonian. 
   We  note that $A_{\alpha}(S, 0)$ is smaller for $ S=3$. 
   For $  \vert\lambda\vert <1$,  this follows from
    the fact that the term $\propto  \lambda ^5 $   has  a large coefficient 
    which changes its sign with the parity of $S$.  
  For $m= -1$, the largest value of     $A_{\alpha}(S, -1)$   occurs for $ S= 3$. We note that 
 $ p(3,-1, \lambda) > 0 $   when $\lambda> 0.7 $. This   indicates  that  the state $ \vert 3,-1\ket $ is   strongly mixed with  the states  $\vert 3, m= 1 ,3 \ket  $  
 within the range $.7 \leq \lambda  \leq 2$.
 
  To conclude this subsection, we emphasize that the holonomic entanglement
  method of $ N>2$ non-correlated   $\frac{1}{2}$-spins, described in reference \cite{bou3}, 
  relies upon the fact that Berry's phase 
has a strong dependence upon $S$.  A different example is given in Appendix A.  In a theoretical note we show how to exploit the difference between Berry's phases for $S=\frac{3}{2}$ and $S=\frac{1}{2}$, with $m=\frac{1}{2}$, to  perform an holonomic entanglement between three initially non-correlated  $\frac{1}{2}$-spins. 
   \subsection{A summary of the non adiabatic corrections to  Berry's cycles }
   To perform empirical determinations of the Berry's phase, one must tackle 
 the problem of the non-adiabatic corrections. This question becomes crucial in the experimental situations where the coherence decay time induced by application of the Hamiltonian $H(\vB,\vE)$ puts constraints on the cycle duration. Shortening the quantum cycle risks to spoil the validity of the adiabatic approximation. 
 In the present context, we have found convenient to study the quantum cycle in the rotating frame attached to the time-varying fields. The Coriolis effect generates an extra magnetic field $ \Delta \vB$ which involves a linear combination of the Euler-angles time derivatives. The longitudinal
component along the $ \vB $ field is the only one which survives when $\alpha$  is the
sole time-dependent Euler angle. The rotating frame Hamiltonian $ \wt{H}_{//}$   is void of any geometry. As a consequence, the phase shift acquired  at the end of
 the cycle $ \tilde{\phi}(\vB + \Delta \vB)$ is purely dynamical.
  The laboratory Berry's phase contribution $\gamma(m)$   is  incorporated into the dynamical phase under the form of its first-order contribution with respect to $ \eta= - \Delta B_{//} /B =
  (\cos {\theta} \; \dot{ \varphi} + \dot{\alpha} )/(\gamma_S B )$. 
 The higher-order terms give all 
 the non-adiabatic corrections associated with $  \dot{\alpha} $ when it is the only
 varying periodic parameter. We have also shown that the subset of these corrections,
 odd under a reversal of $ \eta $, cancel exactly for ``magic" values  
$ \lambda =\lambda^*(\eta)$, obtained from a polynomial fit performed on the numerical results 
valid for $0 \leq \eta \leq 0.5$ (see \cite{bou3} Sec.V.A)
\bea
\lambda^{\star}(2,\eta )&=&0.838213 - 0.0837823 \;\eta^2 - 0.0431478 \; 
\eta^4 - \nonumber\\
&&0.0231887 \; \eta^6 - 
 0.0207986 \; \eta^8. \label{lambdamag}
\eea
This cancellation is implemented in the experimental
project described in Section IV. The $\eta$-even corrections are eliminated by subtracting the phases measured for two ``mirror'' cycles ($\eta \rightarrow -\eta$). The case of the non-adiabatic corrections induced by
the transverse field $\Delta B_{\perp} /B$ is somewhat more involved since it introduces a non-trivial
geometry and, as a consequence, a Berry's phase contribution to be added to the one coming
from the transverse dynamical phase; the results explicited in \cite{bou3} are not used in the present context.   

For the adiabatic approximation to be satisfied, the primary condition is that, 
at each instant, the spin  quantum  state is   an eigenstate of $ H(\vB(t), \vE(t))$. 
In the above discussion, we have made implicitly the  two following assumptions: i) during an  
Euler angle cycle, $\lambda(t)$ has a slowly varying value  of the order of unity, ii) the initial  state of the Euler cycle is obtained by  an adiabatic ramping governed by 
 $ \mathcal{H}( \lambda(t) )= S_z+ \lambda(t)  S_x^2 $, starting from  a null value of $\lambda$  in order to  get its desired value for a well defined value of   $m$. 
 We have shown, by doing explicit calculations, that approaching this ideal ``adiabatic ramping''      is, in practice,  a non-trivial task. The key-parameter which governs the time dependence of the quantum state  is the time derivative of $\lambda$.    
 A linear increase of $\lambda(t) $  would be equivalent  to
a rf pulse with sharp edges, leading to large oscillating non-adiabatic corrections 
 exhibited in ref. \cite{bou3}. Among several methods \cite{guery}, a standard procedure to smooth them 
  out is  to use a Blackman pulse shape \cite{Black}. In the present  context, this condition
  is implemented by taking  for  $\lambda(t) $ the following time dependence:
   $ \lambda(t)=  \lambda_0 f (t/\mathcal{T})/f(0)$,  where $f(s)$ is  the Blackman function
   \be 
  f(s) = 0.42  -0.5 \cos{(2\pi s)} + 0.08 \cos{(4\pi s)} \label{BM}
  \ee
 and  $\mathcal{T}$ the ramping  time. 
  The  efficiency of the procedure has been illustrated before (see Fig. 5 in \cite{bou3}).
  
 There is a third  assumption  implicit in the rotating frame analysis, namely that 
 the adiabatic  approximation is valid for the rotating frame Hamiltonian $\wt{H}_{//}$.
 As will be shown by the theoretical analysis of the experimental project presented in
  section IV, the adiabatic approximation works beautifully provided one uses also
  a Blackman-pulse shape for the angular speed, $ \dot{\alpha}(t) $.    
  
  When looking at a plot of ${\cal E}(m,\lambda)/\lambda$ (Fig.2), it appears clearly that there are pairs of states associated with $m$ values differing by 1 which never cross but become quasi non-degenerate for large values of $\lambda. $ 
When one is interested in quantum cycles where only $\alpha$ and $\lambda$ are  time-dependent this should not affect the results since the Hamiltonian $H(\vB(t), \vE(t))$ has no matrix element connecting the  $\Delta m = \pm1$ states. Nevertheless, when two such levels happen to be close, the spin system becomes particularly sensitive to imperfections which alter the symmetry, as for instance a stray component  of the  magnetic field orthogonal to $\vB$.   

 \section{Atomic   simulation   of  isolated spins non-linearly 
 coupled  to external fields.  }  
 Since, among cold atoms, alkali are the most frequently studied    
we first examine what kind of measurement looks possible in their case.
 Such a  choice is motivated  by the fact these systems can be kept in a 
decoherence-free  space for a relatively long time (up to 1~s), if cooled and trapped. 
But as is well-known, it is very difficult to observe  a static quadratic Stark effect  on 
 hf sub-levels of alkali ground states, since  
their electric tensor polarizabilities   are strongly 
suppressed. A tiny effect appears only as the result of a third-order perturbation \cite
{sand,weis}, when the tensor 
part of the hf interaction is taken into account. The  
 tensor polarizability of Rb $\alpha_2\simeq  $2~mHz/(kV/cm)$^2$ \cite
{gould}, implies  that a field of $\approx$ 300~kV/cm would be required to generate 
a coupling strength of 100~Hz. This makes it unrealistic to use a rotating static E-field to observe the
quadratic Berry's phase. 

On the other hand one may rely on light shifts \cite{cohen1}.  
 Hereafter, we consider the case of a quadratic coupling induced by linearly
polarized light fields which have the advantage, with respect to microwave- or rf-fields, of making the orientation
of $\vE $ relative to $\vB$ both precisely ajustable and easily rotated.   
However, with this method one has to face the problem of instability of the dressed atomic state if  
one wants to obtain a quadrupole to dipole spin coupling ratio of order 1.
 This is not straightforward for the ground state hyperfine levels  of 
alkali atoms. In this case, the role of the spin operator $ \vS $ is played by
the total angular  momentum operator $\hbar \vF = \hbar(\vs +\vI)$,  where $\hbar\vs$ and
$\hbar\vI $ are the electronic and nuclear spin operators. 
Actually, we are going to  show that the practical realization of the  Hamiltonian $H(\vB(t),\vE(t))$  
required to test the theoretical  predictions of this 
work for $\lambda \simeq 1$ appears possible for $^{87}$Rb atoms. 

 Now, in the rapidly expanding 
family of laser cooled and trapped atoms has appeared the 
daring alternative of chromium.
The great progress achieved in the manipulation of this atom have given rise to a series of beautiful
experiments, {\it e.g.} \cite{brad,santos,gorceix}. As a spin $S=3$ candidate, it has the advantage of being of pure electronic origin. This makes the instability of the dressed atom no longer a problem.
The chromium option is  discussed at the end of this section.
\subsection{ Building quadrupole spin couplings  in alkali atoms  using the ac Stark effect.}

 It is known that the application of a light beam close 
to resonance with one
atomic excited state generates light shifts which can simulate the effect of an electric  
field or a magnetic field \cite{cohen1}.  
A fictitious $\vB$ field arises if the beam is circularly polarized while a fictitious 
$\vE$-field is created by a linearly polarized beam. One might wonder therefore, whether, 
with a single beam of elliptically polarized light, it would be possible to 
generate 
both fictitious electric and magnetic fields satisfying the condition $\vE 
\cdot \vB = 0$. 
We consider the case of alkali atoms irradiated by a laser beam nearly resonant 
with one hf component of the $nS_{1/2, F}- nP_{1/2, {\cal F} }$ transition. 
The atom-laser coupling responsible for  ac Stark shifts - or light shifts - of 
the ground state 
sublevel can be calculated to second order in the  
atom-radiation field interaction, in the rotating wave approximation (see \cite{cohen1,cohen2}).  
If we make the simplifying assumption $F={\cal F} $, discussed hereafter, we can use the  proportionality 
between the atomic electric dipole and the angular momentum operators, 
resulting from the 
Wigner-Eckart theorem, $ \vD/ea_0 =d\, g_F \vF$, where $d$ is the $\Delta 
m_s=0 $ 
matrix element of this electric dipole transition in atomic units and
$g_F = 2(F-I)/ (I+1/2)$. In this case, the light shifts of the  hf ground state 
$S_{1/2,F}$ can be 
represented very simply in terms of the effective Hamiltonian
\bea
&&\widehat H_{ls}(F)= \frac{\hbar \Omega^2}{\Delta} g_F^2
\times \nonumber\\ 
&&\hspace{-9mm }\left ( i \hat \epsilon{*} \wedge  \hat \epsilon 
\;  \cdot F\hat k   +
\frac{1}{2} (\vF \cdot \hat \epsilon^{*}  \; \vF \cdot \hat \epsilon  +
\vF \cdot \hat \epsilon \; \vF \cdot \hat \epsilon^{*})  \right)~,
 \eea
where $\Omega = d{\cal E}/2\hbar$ represents the dipolar coupling of the 
atom with the laser field.  
 The field magnitude ${\cal E}$ is related to the photon number density by
$\epsilon_0 {\cal E}^2 = N\hbar \omega/V $ \cite{mab,deut} and the complex vector $\hat \epsilon $ of unit norm defines the polarization. For simplicity, we suppose 
the  detuning $\Delta$ between the laser and the atomic transion frequency,
 $\omega_{F,{\cal F}}$, to be small compared to the hf splittings, 
 so that we can consider the contribution of the 
${\cal F} =F$  hf line alone,  $\Delta= \omega - \omega_{F,{\cal 
F}}$. In the next subsection, we shall consider the case where this condition no longer applies. It will turn
out that the general form of $H_{ls}(F)$ is the same but with different coefficients. 
 
   A general expression for the elliptical polarization
     $\hat \epsilon $ of a beam directed along $\hat z$ is:
\bea
&&\hat \epsilon= \frac{1}{\sqrt{2}}(\cos{\delta_e} 
\;\hat e_{+} + \sin{\delta_e} \;\hat e_{-})  \nonumber \\
«&&  {\rm where} \hspace{5mm} \hat 
e_{\pm}= \frac { \hat x \pm \, i \hat y}{\sqrt {2}} \exp{(\mp \; i u)},
\eea
and the angle $u$ defines the orientation of the ellipse in the $(x,y)$ plane of polarization.  

For $u=0$ the expression for $\widehat H_{ls}$ obtained after some 
angular momentum algebra is:
\bea
&&\widehat H_{ls}(F) =  \frac{\hbar \Omega^2}{\Delta} 
g_F^2 \times\nonumber \\
&&\hspace{-5mm}\left ( \cos{2 \delta_e}\; F_z + \sin{2 \delta_e} \;F_x^2
+ (F(F+1)-F_z^2) \frac{1-\sin{2\delta_e}}{2}\right ) \nonumber\\\label{Hls}
\eea
The first term, linear in $\vF$ involves the circular polarization of the beam, 
characterized by its helicity $\xi= {\rm Im}\{\hat
\epsilon^{*}
\wedge \hat \epsilon
\cdot
\hat k\}= \cos{2 \delta_e}$, while the quadratic contribution $\propto F_x^2$ involves 
the linearly polarized intensity.
Contrary to a static $\vE$-field, a laser field can provide a sufficiently  strong quadrupolar 
coupling, for laser power and frequency adjustments lying within  
convenient limits (see Sec IV). 
However, because the ac-field is complex, while the dc field is real, 
$\widehat H_{ls}(F)$ actually 
differs from the Hamiltonian $ \widehat H(B,E)$ of Eq.(\ref{themodel})
by its last term involving $F_z^2$, which  cancels out only if the polarization 
is purely linear, {\it i.e.} $\delta_e = \pi/4.$ Although this does not render the light-shift 
Hamiltonian untractable, one looses the  simple  symmetry property of the Hamiltonian
 under  the rotation  ${ \cal R}(\hat x ,\pi) $ discussed   in section II.B. 
For this reason, to keep our concrete discussion of the theoretical 
implications as simple as possible, we prefer to choose
 an experimental  situation corresponding to
{\it a hybrid realization} of the field configuration. We shall suppose that
$\vB= B \hat b$ is a static field, while $\vE= {\cal E} \hat e$ is a light field, linearly polarized along the 
direction $\hat e$, the light beam direction $\hat k$ being taken parallel to 
$\hat b$.   With this hybrid ``$\vB$-field light-field''
Hamiltonian, one can  satisfy exactly all the conditions required in ref. \cite{bou3}, 
by allowing $\hat e$ to rotate around 
$\hat b \parallel \hat k$ at the angular speed $\dot{\alpha}$. This hybrid Hamiltonian   
\be
H_{hyb}(F,t) = \gamma_F\,\hbar \, B \; \vF \cdot \hat b 
+\frac{\hbar \Omega^2}{\Delta} g_F^2  (\vF\cdot \hat e)^2,
\label{Hhybrid}
\ee
 is identical to the Hamiltonian of Eq.(1), if one makes the 
correspondences   
\be
    \gamma_Q E^2 \hbar ^2\leftrightarrow  \frac{\hbar 
\Omega^2}{\Delta} g_F^2  \hspace{5mm} {\rm and}
\hspace{5mm} \lambda \leftrightarrow  \frac{\Omega^2}{\Delta} g_F^2 / 
\gamma_F B.\label{ls-lambda}
\ee
We set  $\gamma_F B = g_F \gamma_s B$, where $\gamma_s B$ is the Larmor 
angular frequency of the electron and $g_F= 2(F-I)/(2I+1)$.  
(Of course the fact that the internal angular momentum $\vF$, relative
to  a given hyperfine sub-level, can be treated  as an isolated spin implies the reasonable assumption that, for values 
of  $\lambda \sim 1$,  the  Larmor  frequency is much smaller than  hyperfine splitting.)
The use of an ac-light field instead of a static electric field has other advantages besides the
magnitude of the Stark coupling. By adjusting the laser detuning, it makes it possible to apply a light field upon a single  ground state hf level,  the second one remaining spectator and providing the phase reference.
The expression for
$\lambda$ also shows there are two independent ways of reversing the sign of the Stark {\it versus} the Zeeman-coupling: one can reverse either the sign of $\vB$  or else that of the laser detuning.

\subsection{ Solving  physical problems raised  by the instability  of  the ``dressed" 
atomic  ground state. } 
 
The main drawback of the light-shift method for obtaining an effective $\vE$ field, is that it generates 
an instability of the ``dressed"  ground state  hf sublevel, which is going to simulate our isolated
spin system. This instability   is best understood within a fully quantized  description  of the atom 
interacting with the radiation field. The vector  state of the ``dressed"  atom of interest 
can be written as: $ \vert F \,nS_{1/2}\ket   \otimes  \vert N\, \hat \epsilon \, \omega \ket $
where $ N $ is the number of photons with energy $ \hbar \omega $ and linear polarization $ \hat \epsilon $
in  the coherent light beam. The light shift  is associated with the two  successive virtual electromagnetic transitions:
$ \vert F \,nS_{1/2}\ket   \otimes  \vert N\, \hat \epsilon \, \omega \ket  \Rightarrow 
  \vert  \mathcal{F} \,nP_{1/2}\ket   \otimes  \vert N-1\, \hat \epsilon \, \omega \ket  \,   \Rightarrow  
     \vert F \,nS_{1/2}\ket   \otimes  \vert N\, \hat \epsilon \, \omega \ket $. More precisely
   $\vert  \mathcal{F} \,nP_{1/2}\ket $,  for an appropriate  detuning,  is   returning 
  by   stimulated emission  to the same ground state hf sublevel   $\vert F \,nS_{1/2}\ket $  but with   its energy ``light-shifted".
 An alternative route for 
  the  excited state $\vert  \mathcal{F} \,nP_{1/2 }\ket $ is  to  make transitions to 
       ground state  hf sublevels,  by an energy conserving spontaneous light  emission,
        leading to the infinite  set of  final states:
    $  \vert F^{\prime}   \,nS_{1/2}\ket   \otimes 
     \vert N-1\, \hat \epsilon \, \omega ; 1  \;  \hat \epsilon^{\prime}  \,  \omega^{\prime}   \ket  $ where  
    $\omega^{\prime}  \neq \omega$ and $ \hat\epsilon^{\prime}  \neq \ \hat \epsilon $. We see
    easily  that all these   final states are orthogonal  to the initial ``dressed"  ground state  hf sublevel even if $ F= F^{\prime}$. In conclusion, our candidate for an isolated spin system is unstable 
    with a decay rate $\Gamma_{dec} $ which scales  as 
    $(\Omega ^2/ \Delta^2) \Gamma_{\,nP_{1/2}}$,   where $\Gamma_{\,nP_{1/2}}$ 
    denotes the spontaneous emission rate of the excited state $nP_{1/2}$.   
 
 At first sight, since the light-shift involves the inverse of the detuning, while the decay rate scales 
as the square of this quantity, the choice of large detunings would seem the most
appropriate. In fact, this is not so.  Indeed,   for $\hbar\Delta $ much larger than the hf frequency
splitting of the excited $P_{1/2}$ state,
$\Delta {\cal W}_P$, the two hf states contribute with nearly equal magnitudes but
opposite  signs, in such a case the light-induced quadratic spin-coupling vanishes. 
 Therefore, the quadratic coupling is easier to achieve at detunings comparable to $\Delta{\cal W}_P$ for heavy alkali atoms (Rb, Cs) or odd isotopes of  
alkali-like ions (Ba$^{+}$, Hg$^{+}$), which have the largest hf splittings.  
Since the $P_{3/2}$ hf splitting is smaller than that of $P_{1/2}$ by a factor 5 in Rb, the light beam
should be tuned preferably close to the $D_1$ resonance line. Detunings of the order of this
splitting (0.816~GHz for
$^{87}$Rb) appear to lead to a reasonable compromise solution between decay rate and light shift magnitude (comparable to the  Zeeman coupling for a magnetic field in the 0.1-1 mG range, see Sec. IV).      
For a correct  evaluation of  $\lambda$, we have to include 
 the  contributions from  both hf  $ nP_{1/2}$ levels,  with their respective detunings  $\Delta_{F,{\cal F}}= \omega - \omega_{F,{\cal F}}$.
The explicit  formula giving   $\widehat H_{ls}(F) $ in the case of an alkali ground state  hf level $F$ ($g_F=1/(I+1/2)$), lightened by a linearly polarized  light beam becomes \cite{mab,deut} 
 \bea
 H_{ls}(F)&= &\frac{\hbar \Omega^2}{\Delta_{2,1} +i\Gamma_P/2}(\mathbf{1} - g_F^2 (\vF \cdot \hat  e )^2 ) \nonumber\\
  &+ &\frac{\hbar \Omega^2}{\Delta_{2,2    
} +i\Gamma_P/2} g_F^2 (\vF \cdot \hat  e )^2 \label{ls2}.
\eea
A remarkable feature of this expression is that, as announced, the two  hf  $ nP_{1/2}$ levels 
  quadrupole  contributions cancel out if the frequency detuning is much larger than the 
$nP_{{1}\over{2}}$ hf splitting, $ \Delta_{2,2}\simeq \Delta_{2,1} $.
In  the sum with equal weigths of the hf levels  numerators  the nuclear spin dependence disappears; only remains
the effective transition dipole $ \propto   \vs $.
The instability discussed above has  been accounted for 
by adding to the energy denominator  $+i\Gamma_P/2 $, see ref.\cite{cohen2}.

Ignoring for a  moment  the natural  width compared to the  detuning, the quadrupole  
coupling term  is given by an expression identical to Eq. (8).  With the definitions:
\be
W_{ls} = \hbar   \frac{
\Omega^2}{\bar\Delta}g_F^2  \hspace{5mm} {\rm and} \hspace{5mm}
1/\bar{\Delta} = 1/\Delta_{2,2}-  1/\Delta_{2,1}  \label
{lambhyb}. 
\ee
we recover the expression of the ``B-field light-shift'' Hamiltonian:
\be
H_{hyb}  = Ê\hbar  \gamma_F\,  B \; \vF \cdot \hat b 
+ W_{ls}  (\vF\cdot \hat e)^2.  
\label{lambhyb}
\ee
deduced from Eq.(\ref{Hhybrid}) by substituting $\bar \Delta$
in place of  $\Delta$. 

 Let us now turn to  the  physical effects induced by   the imaginary part
 of the energy   denominators. 
  The most dangerous effect involves 
 the imaginary part of the quadrupolar coupling  given by the  expression  
$$g_F^2\frac {\Gamma_P}{2} \Omega^2 (\vF\cdot \hat e)^2 ( \frac {1}{\Delta_{2,1}^2}-\frac {1}{\Delta_{2,2}^2}),  $$
up to third order in $\Gamma_P/ \Delta$. 
This contribution modifies during the quantum cycle the structure  of the atomic wave function, which is a linear  combination of m-dependent angular momentum states, and hence this   
invalidates the  Berry's phase derivation of   
 ref. \cite{bou3}. There is, fortunately, a remedy to this problem.
  It is to tune the laser frequency  midway  between resonance with the two hf states as illustrated in the insert of Fig.(\ref{fig3}). Then, the imaginary part of $ \lambda $  is vanishing. From the remaining scalar part results a decay rate $\Gamma_{dec}$ of the dressed atoms, 
   \bea 
&& \hspace{-7mm}\Delta_{2,1}^2= \Delta_{2,2}^2  \;\text{ implying }\; \vert\Delta_{2,2}\vert=\vert\Delta_{2,1}\vert = 2 \pi \Delta {\cal W}_P/2, \label{Delta}\;\;\\
&&{\rm and}\;\; W_{ls}=  \frac {4 g_F^2\Omega^2}{2\pi \Delta {\cal W}_P },
\;\;\; \Gamma_{dec}=\frac{ 4\Omega^2}{(2 \pi \Delta {\cal W}_P)^2} \Gamma_P . \label{gammasurW}
\eea
From now on, therefore, we shall assume that the laser beam frequency detuning satisfies condition
(\ref{Delta}) and we  rely on Eqs.(\ref{lambhyb}) and (\ref{gammasurW}), to suggest experiments on
 alkalis.  We are left with an ``isolated"  spin with decay rate  $\Gamma_{dec}$.
For dressed  $^{87}$Rb atoms, the spectroscopic parameter governing  the instability turns 
out to be small:
\be
\frac{\Gamma_P}{2 \pi \Delta {\cal W}_P}= 7 \times 10^{-3}, \label{WPoverGama}
\ee
instead of 6 and 5 $\times 10^{-2}$ for $^7$Li and $^{23}$Na.
As shown in Sec. IV, $^{87}$Rb thus provides a unique possibility to test Berry's cycles for $S=2$ in the range $0<\lambda \leq 1$. 

 In the last section we shall be interested in the case of a spin-1 performing a quantum cycle not in the Hamiltonian parameter space but {\it in the density matrix space } described by the atomic polarization and alignment. The hybrid light-field B-field Hamiltonian considered in this section is adapted to realize a cycle of this kind. The laser beam in this case will be equally detuned from the two
 hf lines starting from the $F=1$ ground state, therefore $\Delta_{12}^2=\Delta_{11}^2$. It is easily verified that all preceding equations  remain valid if $\Delta_{12}$ is changed into $\Delta_{21}$ and $\Delta_{22}$ into $\Delta_{11}$. We shall need the explicit hybrid Hamiltonian for $F=1$, in the case of an elliptically polarized laser beam, with ellipticity axes at 45$^o$ of the $x$ and $y$ axes, {\it i.e. $u=\pi/4$}, and a B-field along the beam:  
 \bea
 H(W_{ls}, B, \delta_e)/\hbar= W_{ls}\(( \frac{1}{2} \sin (2\delta_e)  \{F_x F_y\} - \frac{1}{2} F_z^2 \)) \nonumber \\
  +  ( W_{ls}\, \cos (2 \delta_e)  + \gamma_F B )  F_z + W_{ls} \,\frac{\vF^2}{2}, 
  \label{Hhybrid-delta}\\
{\rm with} \;\;\;  W_{ls} = \frac{ \Omega^2}{2\pi \Delta {\cal W}_P }. \label{Wls} \hspace{3.5cm}
   \eea   
 \subsection{The special case of $^{52}$Cr chromium isotopes with S=3}
   Chromium atoms in their ground state have also $L=0$ like alkalis,  but they possess
a large electronic spin $S=3$ and an isotope of 84$\%$ natural abundace,  $^{52}$Cr,
without nuclear spin.
The ground state $^{7}S_3$ is coupled by dipole transitions to 
different P-states ($^{7}P_{2,3,4}$) which have been used for cooling the atoms 
in an optical dipole trap \cite{brad}. With
linearly polarized blue lasers suitably detuned from these 
transitions it is possible to induce Stark
shifts proportional to $m^2$ in the ground state. For instance in the 
particular case of a small detuning
with respect to the $J$ conserving transition $^7S_3 \rightarrow ~  ^7P_3$,  
equation (\ref{Hhybrid}) could be adapted by replacing $\vF$ by $\vJ$ and performing some angular momentum algebra. Like in the case of alkalis,
detuning and light power have to be adjusted to minimize 
the instability of the dressed atomic state. 
However, as in chromium the fine 
structure of the P-states
being about 300 times larger than the hf structure of the alkali 
P-states, those restrictions are much easier to
satisfy. Importantly, it is also much easier to prevent the quadratic coupling effect 
from being affected by the instability.   
This quadratic Stark shift has been observed, and exploited for the study of
spin-3 Bose-Einstein condensates in the geometry 
$\vE \parallel \vB$  \cite{santos} (where it was termed ``quadratic 
Zeeman effect''), but we are interested here in the geometric phases in the configuration $\vE \perp \vB$.  
 
\section{ Possible atomic interferometer measurement of Berry's phases
for spins with quadrupole  coupling  }
 Among several possible tests of the most striking results 
associated with the quadratic Stark coupling, there is an important goal which is to first 
observe the Berry's phase generated by the rotation of  $\vE$ around $\vB$. 
\be  
\beta(m=0, \lambda)= - \oint  p(0, \lambda) d \alpha.
\ee
This is one of the remarkable effects signalled out in Sec. II, that we expect to be large for $S=2$ when $\lambda$ is close to 1, as highlighted by figure 1.  
The effect is particularly noteworthy for  
a spin larger than one, initially in an $\vert S,0\ket $ substate,
{\it i.e.}  having its quantum
averaged polarization along $\vB $ cancelling out before an adiabatic
cycle starts and after it stops. Therefore, the  purpose of this section is to make precise suggestions for a  measurement  involving the rotation of $\vE$ around the orthogonal $\vB$ field, when both couplings, linear and quadratic, are of comparable magnitudes.  
\subsection{Experimental compromises for spin-2 measurements with matter-wave 
interferometers. } 
Measurement of the phase difference acquired during the
evolution of a quantum state requires a phase reference. Therefore the methods of matter-waves
interferometry,  first employed with  neutron beams 
\cite{Bon,bitter}, then adapted to
magnetic resonance
\cite{tycko,pines} and now the subject of  outstanding developments in
cold atom physics \cite{interf}, are
especially well suited to observe Berry's phases. Atomic interferometry has been used 
before to measure a topological phase \cite{Hin,Zei,Weis,Ska,Foot}, but not in 
the case of a quadratic spin coupling, where the parameter space 
cannot be reduced to the surface of a sphere.

The initial atomic state is represented by a coherent superposition of two states having 
different energies, such as for instance two hf substates of an alkali atom. The main 
difficulty arises from the decay rate $\Gamma_{dec}=\frac{
\Omega^2}{\Delta_{2,1}^2} \Gamma_P $
  (see section III.B   Eq.(\ref{gammasurW})). 
   For concreteness,  
  we shall illustrate our proposal on the
case of the two hf substates
of $^{87}$Rb, $\vert F=1,m=0\ket $ and $\vert F=2,m=0\ket $.
These have been chosen because first, only one of these two mixed
substates acquires a Berry's phase; secondly, as we shall show, the range of parameters
leading to $\lambda
\approx 1$ looks achievable for this isotope.
By  making a specific choice of the experimental 
  parameters, we have found that an interferometer measurement 
  of the peculiar features of the $ S=2,\, m=0$ Berry's phase is feasible 
 within the present state of the art.
 However, as we shall see, there is actually little freedom for 
organizing the quantum cycle,  if one wants
to perform a measurement,  free of 
non-adiabatic corrections.  

i) {\it Instability versus adiabatic requirements} 

The relevant range of the $\lambda$ parameters ($ 0<\lambda\lesssim 
1.5 $), has to be reached with moderate laser intensities, in order to limit 
the decay rate of the dressed atoms. It requires  the
magnetic field  to be small ($\simeq 1$ mG), but not  much smaller, in order 
to keep the  field homogeneous  over the atomic sample.
The problem of the dressed atom losses would favour rapid quantum cycles
but  they should  not spoil  the validity  the adiabatic approximation, 
 in  particular  during the preparation of the 
coherent quantum state before it is submitted to the rotating 
electric  field.
 It is then   crucial  to tame out  the non-adiabatic oscillations  
 generated by a linear ramping of $ \lambda(t) $ from zero to a value $\simeq 1$.
 An efficient remedy, given  in \cite{bou3},  is  to use a Blackman pulse-shape
for  the time derivative $ \dot \lambda(t)$ (See II.D)
 
 As regards to the non-adiabatic correction associated with the angular
velocity of the $ \vE$ field rotation, 
they are governed by the  convenient  parameter $\eta= \dot \alpha /\gamma_s B.$
We  have shown in  our previous  work   \cite{bou3} that all the corrections to the Berry's phase
of the order $ \eta ^{2 n+1}$ vanish for  $n \geq 1$,  if 
 $ \lambda(t) $  is given by the magic value $\lambda^{\star}(2 ,\eta(t) )$ ( See Section IID.)
  We shall see that the set of values $\eta = 0.3$ 
and $\lambda^{\star}(\eta) =
0.830$,  is an element  of the overall  compromise  to be  elaborated in the next 
subsection. 

ii) {\it Absence of quadrupolar light shift in the $F=1, m=0$ quantum state}.

The cancellation of the quadratic light shift in the $\vert 1,0 
\ket$ quantum state of the
$^{87}$Rb atom results from its absence of $m = \pm 2 $ sublevels. 
This simplifies the
preparation of the coherent state which becomes a superposition of 
this unperturbed substate with
the state perturbed by the laser field. This latter can be written
$\vert \Psi(2,0; t=0)\ket\equiv \vert
\hat\psi(2,0;\lambda(0))\ket$ in the  initial state of the quantum 
cycle; and is thus an
eigenstate of $
\hat H(\vB(t), \vE(t))$ defined by
  Eq. (\ref{lambhyb}). (Note that it is now useful to specify the spin 
value, since we are dealing with an admixture of two different hf states, 
hence belonging to different internal spin spaces).
  \begin{table*}[t]
\caption{ \small Experimental parameters for a Berry's phase 
measurement with $^{87}$Rb
atoms supposing $\Delta_{2,1}=-\Delta_{2,2}$, and using  time-independent  
``magic'' conditions
$\eta= 0.3$,  $\lambda^{\star}(\eta) = 0.83030$. Values are given 
for two different laser
intensities $I$ expressed in terms of the ``saturation'' intensity $I_{sat}$, defined as 
$\Omega_{sat}^2=\Gamma_P^2/8$.  For the $D_1$ line of Rb 
$I_{sat}\approx $ 3.3 mW/cm$^2$. The product $\Gamma_{dec}T_c$ does not depend on $I$ (see the text 
and  Eq.(\ref{nodec})). }
\begin{center}
\begin{tabular}{p{1.5 cm} p{1.5cm} 
p{1.5cm}p{1.5cm}p{1.8cm}p{1.6cm}p{2.0cm}p{1.8cm}p{1.5cm}}
\hline\hline
$\small\frac{\Delta_{2,1}}{ 2\pi \Delta {\cal
W}_P}$&$\small\frac{\Delta_{2,2}}{ 2\pi \Delta {\cal W}_P}$ &
$\small\frac{\bar{\Delta} }{ 2\pi \Delta {\cal W}_P}$
   & $~~\small{\Gamma_{dec}T_c}$    & $\small{100\frac{\Omega^2}{\Gamma_P^2}}$
  & $ I/I_{sat}$ & $\small{\gamma_F B^{\star} {\scriptsize\text(\text 
{s}^{-1})}}$ & $~\scriptsize{
~{\Gamma_{dec}(\text{s}^{-1})}}$&$\scriptsize{~~T_c(\text{ms})}$
  \\
\hline
  &&&&&& &~~\\
\vspace{-5mm}
~~~0.5 &\vspace{-5mm}~~~{-0.5}  &\vspace{-5mm}
~~-0.25 &\vspace{-5mm}~~~0.244 &\vspace{-5mm} ~~~~1.4 ~ &\vspace{-5mm} 0.113
&\vspace{-5mm} ~~~4400  &\vspace{-5mm} ~{~101}&\vspace{-5mm} ~~~~2.4 
\\ \hline
  &&&&&& ~~\\
\vspace{-5mm}
~~~0.5 &\vspace{-5mm}~~~{-0.5}  &\vspace{-5mm}
~~-0.25 &\vspace{-5mm}~~~0.244 &\vspace{-5mm}~~~~0.14  &\vspace{-5mm} 0.011
&\vspace{-5mm} ~~~~~440  &\vspace{-5mm} ~{~10.1}&\vspace{-5mm} ~~~24   \\
  \hline \hline
\end{tabular}
\end{center}
\end{table*}
  \begin{figure*}
  \vspace{10mm}
  \centering
 \includegraphics[height=7 cm ]{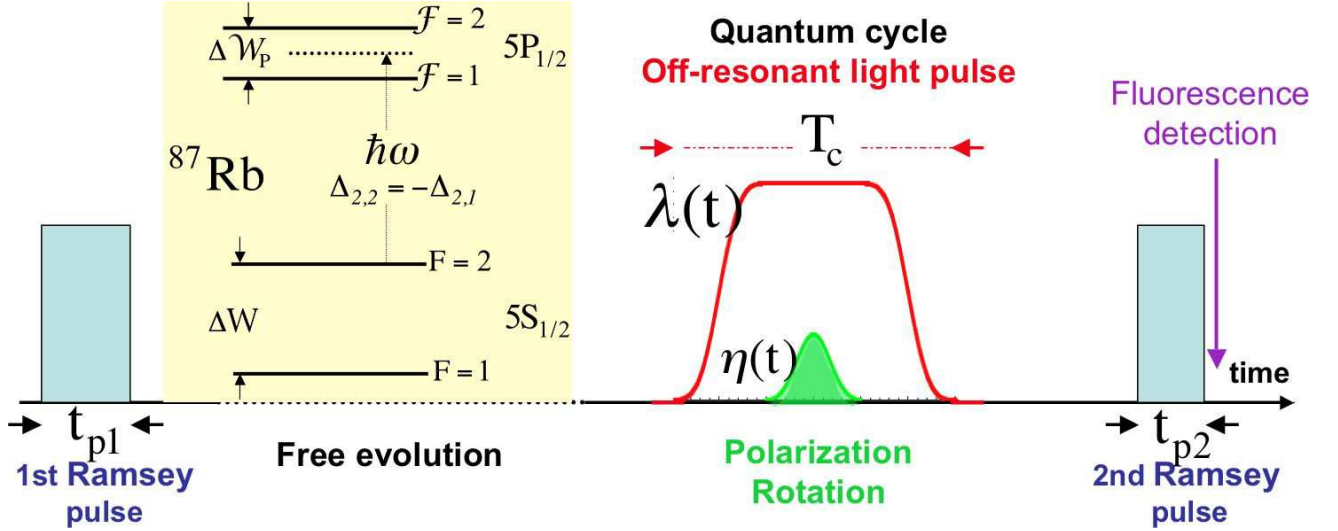}
  \caption{ \small (Color online) Interferometric measurement of the Berry's phase 
for a spin 2 ($m=0$), chronology of the interferometric cycle 
including the quantum cycles, time in arbitrary units.
The two phase-coherent Ramsey pulses are resonant for the $1,0
\rightarrow 2,0$ transition of unperturbed $^{87}$Rb atoms. The
quantum cycle of atoms in the state  $\vert 2,0 \ket$ starts with a 
ramping-up of the intensity of the
off-resonant  laser field, then, the rotation of its
linear polarization starts smoothly towards a maximum angular speed.
The whole operation is followed by the time reversed one to
return to the initial state. The temporal dependences of both the
light shift - normalized by the Zeeman shift, $\lambda(t)$- and the
angular velocity of the polarization rotation - normalized by the
Larmor precession angular frequency, $\eta(t)$ - as well as their
relative magnitudes are  the parameters which govern the 
non-adiabatic corrections to the
Berry's phase in a critical way. By using Blackman pulses for both 
$\dot \lambda(t)$
and $\eta(t)$ (``oscillation-taming'') and by 
satisfying the magic relation
$\lambda(t) = \lambda^{\star}(\eta(t))$,
we show (see the text) how it is possible to keep these corrections well
below the $0.1\%$ level, with 40$\% $ of the initial cold-atom cloud
surviving the light-induced decay.
Insert: relevant atomic levels. }
\label{fig3}
\end{figure*}
iii) {\it Parameter  compromise for a simplified $ S=2$, $ m=0$ Berry's cycle
performed on $^{87}$Rb: one example.}

Up to the end of this subsection and only here, we shall
analyse the  simplified $ S=2,\;m=0$ Berry's cycle defined 
 by  the boundary conditions: $ \lambda(0)= \lambda(T_c)$ and
$ \alpha(T_c)- \alpha(0) =\pi $ and made the further assumptions $ \dot\lambda(t)=0$
and $ \alpha(t)= \pi/T_c $.  

  Table I presents the relevant parameters for $^{87}$Rb.
For $\Delta_{2,1}= \frac{1}{2}\Delta {\cal
W}_P/\hbar = 2 \pi \times 0.408 $~GHz and $\Omega =2\pi \times 0.68$~MHz
corresponding to a laser intensity of  $\approx$ 0.37 mW/cm$^2$, namely   
0.113 times the ``saturation'' intensity, $I_{sat}$ defined such that $\Omega_{sat}^2= \Gamma_P^2/8$; 
the quadrupolar Stark coupling, $ \Omega^2 g_F^2/ \bar{\Delta} $,
amounts to $2\pi \times $ 575~Hz, while the linear Zeeman coupling is
$2\pi \times $ 700~Hz/mG.  In order to explore the interesting
domain $0.1<\lambda \lesssim 1.5 $, the applied magnetic field should 
lie in the range $ 10 \gtrsim B
\gtrsim 0.65 $~mG. 

For constant angular velocity, the time needed to perform one quantum 
cycle is $T_c= \pi/\dot \alpha = \pi/ \eta \gamma_F B $. 
If we want to keep the signal loss per cycle arising from the light-induced Stark coupling smaller than 1,  the condition to be fulfiled is 
$\Gamma_{dec}\, T_c \lesssim 1 $, or
\be
\Gamma_{dec}\, T_c =\frac{\pi}{g_F^2} \frac{\lambda}{\eta} 
\;\frac{\bar{\Delta}\;  \Gamma_P}{\Delta^2_{2,1}}= 4\pi  \frac{\lambda}{\eta}\;(\frac{\Gamma_{P}}{ 2 \pi \Delta{\cal W}_P }) \lesssim 1, 
\label{nodec}
\ee
using Eq.(\ref{ls-lambda} and \ref{gammasurW}).
{ \it Although the laser intensity is involved in both the expressions for $T_c$ and
$\Gamma_{dec}$,  it disappears from} $\Gamma_{dec}T_c$, but the intensity selected determines the magnetic field range and the minimum time needed for a measurement, which are interrelated together: the higher 
the field, the shorter the time. 
As noted before $^{87}$Rb is among alkali atoms the most favourable one
with $\Gamma_P/ 2 \pi \Delta{\cal W}_P = 7\times 10^{-3}$.
Therefore, one can only select the value of the
ratio  $\lambda/\eta $,   preferably not exceeding a few units,
  for condition (\ref {nodec}) to be satisfied.
    If one wishes to take advantage of the ``magic'' value property of 
$\lambda$, one is led to choosing a rotation speed of the $\vE$ field moderately large, 
($\eta=0.3$, 
 $\lambda^{\star}(\eta)/\eta= 2.76$ for 
$^{87}$Rb). Finally, this choice of experimental conditions (see Table I)  leads to an acceptable 
signal loss for a single  quantum
cycle, $\simeq 24\%$.

   In the next subsection, we propose a realistic precise timing for
the quantum cycle including now a discussion of the atomic loss during ramping up and down of the $\lambda(t)$ parameter. This operation is of a critical importance to avoid 
non-adiabatic corrections. 
  \subsection{Towards an empirical determination of Berry's phases, 
free of non-adiabatic corrections to the few 0.1 $\%$ level}
\begin{figure}
  \centering
\includegraphics[height=4.0 cm]{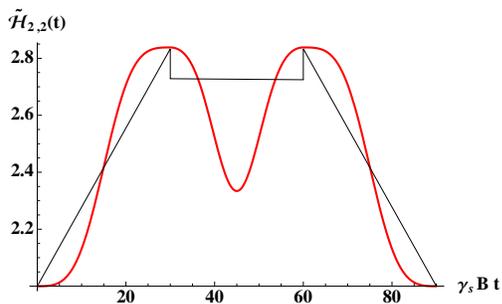}
\caption{ \small (Color online) Time dependence of the diagonal $(S=2, m=2)$ matrix 
element of the Hamiltonian in the rotating frame,  during the whole 
quantum cycle (time unit: $(\gamma_S B)^{-1})$. Black curve: linear 
ramping for $\lambda(t)$ and square pulse for $\alpha(t)$. Red curve: 
taming of the non-adiabatic oscillations (OT procedure) by using 
Blackman pulses for both $\dot{\lambda}(t)$ and $\dot{\alpha}(t)$.}
\end{figure}
 \subsubsection{Organization of the interferometric cycle}
   The interferometric cycle can be organized according to the method 
developed for high accuracy
measurements with hfs measurements in cold atom atomic clocks to prepare the 
unperturbed coherent state
$\propto   (\vert 1,0\ket + c \;\vert  {2 ,0}\ket)  $ ({\it e.g.} \cite{bize}).
Starting from the pure $\vert 1,0\ket$ state in presence of $\vB$
{ \it and without} the laser beam, hence without $\vE$, one can use 
the Ramsey method \cite{ram} to prepare
the unperturbed coherent state
$\propto   (\vert 1,0\ket + c \;\vert  {2 ,0}\ket)  $, {\it i.e.}
apply suddenly a short pulse (using either a rf field resonant for 
the $\vert1,0 \ket
\rightarrow \vert 2,0\ket $ frequency or two
Raman pulses \cite{Kachu}). After a certain delay (time of free evolution, $T$,
long compared to the durations of the rf pulses and the quantum 
cycle), the quantum cycle starts.
The laser field responsible for $\vE$ is applied progressively and 
followed later on by the
polarization rotation. 
  It lasts for a duration $T_c$. Then, detection takes place. This 
consists in the application of a
second Ramsey pulse,  in presence of the $\vB$ field and without 
$\vE$. The Berry's phase is
obtained from the probability of finding the atoms in the $ \vert 
1,0\ket$ (and $ \vert 2,0\ket$)
state at the end of this pulse. The latter is measured by 
fluorescence detection following resonant
excitation of the $5S_{1/2,F=1}-5P_{3/2,F=0}$ (and 
$5S_{1/2,F=2}-5P_{3/2,F=3}$) transitions. The
cycle is depicted in Fig. \ref{fig3}
 \subsubsection{ Implementing the quantum cycle via\\ an 
oscillation-taming procedure}
 \begin{table*}[t]
\caption{Chronology of Berry's cycle organized in three equal steps using $1/ \gamma_S B $ as time unit. The function $h(t)$ is the primitive
of the Blackman function $f(s)$ (Eq.(\ref{BM})), varying between 0 and 1 over the unit time interval, $ h(t) = \int_0^t
f(s) ds / \int_0^1 f(s) ds $.}
\begin{center}
\begin{tabular}{p{4.0 cm} p{4.5cm} p{4cm}p{3cm}}
\hline\hline
  Time interval& $~~~~~~~~~~{\wt{\cal H}(t)} $ &
$~~~~~~~~~~~~ \lambda (t)
$ & $~~~~\eta (t)=\dot{\alpha} (t)/\gamma_sB   $
 \\
\hline
  && ~~\\
\vspace{-5mm}step i) $\; \;\; 0\leq t \leq {\cal T}$ &\vspace{-5mm} $~~~~S_z + \lambda(t) 
S_x^2$&\vspace{-5mm}$~~~~~~~~\lambda^{\star}(0)\;  h (t/{\cal T})$&\vspace{-5mm}~~~~~~~~~~~~~0
  \\
\hline
&& ~~\\
\vspace{-5mm}
 step ii)  $\; \; {\cal T}\leq t \leq  2{\cal T}$&\vspace{-5mm}$ (1 -  \eta(t) ) S_z + 
\lambda^{\star}(\eta(t)) S_x^2 $&\vspace{-5mm} $~~~~~~\lambda^{\star}(\eta(t))$ Eq.(\ref{lambdamag})
&\vspace{-5mm} $~~~~~\frac{\pi}{ \gamma_S B {\cal 
T}} \, \dot{h} \((\frac{t - {\cal T}}{{\cal 
T}}\))$ 
  \\
\hline
&& ~~\\
\vspace{-5mm}step iii) $ 2{\cal T }\leq t \leq T_c = 3 {\cal T }$
&\vspace{-5mm} $ ~~~~~ S_z + \lambda(t) S_x^2 $
&\vspace{-5mm}$\;\;\;\lambda^{\star}(0)\;  h\((( T_c-t)/
{\cal T}\))$ &\vspace{-5mm}~~~~~~~~~~~~~0   \\ 
 \hline \hline
\end{tabular}
\end{center}
\end{table*}
  For clarity we suggest splitting the quantum cycle into three steps 
in which only one parameter at a
time is varied. During the first and the last
steps, $\lambda$ alone is varied from 0 to $\lambda_0$, while during the
second step $\alpha $ is varied from 0 to $\pi $. Moreover, we decide
to comply with the recommendation given and justified in \cite{bou3}
to satisfy the adiabatic approximation, for
preparing the quantum state
$\Psi(2,\,0;\lambda_0)$ at the beginning of step ii) with its 
polarization $\langle S_z \rangle$ very
close to its desired adiabatic value, $p(2\,0;\lambda_0)$. We avoid the
discontinuities in the variation of  $\dot{\lambda}(t)$ by suitable 
tailoring of its shape:
$\lambda(t)$ will be assumed to be described by the primitive of a  Blackman pulse. 
During step 2) the time 
variation of $\dot{\alpha} $ is supposed to be represented by the Blackman
function $f(s)$ (Eq. \ref{BM}); this is what we have termed the oscillation taming (OT)-procedure.
In addition, it is advantageous and possible to adjust $\lambda(t)$ 
so that at any time
it coincides with its ``magic value'' associated with the angular
velocity $\eta(t)$ at that time. This means that during steps i) and
iii) $\lambda_0$ is chosen equal to $ \lambda^{\star}(0)$, while
during step ii), instead of keeping $\lambda$ constant, we make a 
fine tuning of $\lambda$
in order to satisfy Eq.(\ref{lambdamag}). In
practice, the variation of $\lambda^{\star}(\eta)$ in the range 
$0\leq\eta\leq 0.3$ is only
half a percent of $\lambda_0$. The
advantage of this strategy is the complete suppression of 
non-adiabatic corrections to
the Berry's phase, which are odd under reversal of $ \dot{\alpha} $ 
(Sec.V.A), the even ones
being suppressed by subtracting measurements for mirror-image cycles 
($\dot{\alpha}(t)
\rightarrow - \dot{\alpha}(t) $).

 Table II summarizes the chronology of Berry's cycle organized in three steps.  For each step we give the expression the Hamiltonian $ \wt {\cal H}(t)$  acting  within the frame
   rotating around the z axis with the angular velocity $ \dot {\alpha } $, together with the time dependence of the parameters $\lambda(t)$ and $\eta (t)$.  
\subsubsection{Quantitative predictions}
     Choosing $T_c= 3 {\cal T}$ which implies for the angular speed at its maximum $\eta^{max}= 0.27$ ({\it } quite close to the value considered in Table I), we have simulated the exact time evolution.  We have solved numerically the
Schr\"odinger equation for the Hamiltonian represented by the sum of
the three  time dependent  operators defined on the three successive 
time intervals.
(Rounding-off theta functions  are  used to avoid
discontinuities of higher order derivatives at the passage between 
two time-intervals.) The
initial state at $t=0$  is  the pure state $\vert 2,0\ket$. If  our
attempt  to create    the conditions  for  the adiabatic 
approximation is working,   we expect
$\langle S_z \rangle$   to be very close to its adiabatic value at the
end of step i) and at the end of step ii).  The calculation yields
a difference of only 1 part in  $10^{3}$. This indicates that, as 
anticipated from the discussion given in \cite{bou3} (Sec.V.C),
our  implementation of  the Berry's cycle  leads to very small
deviations  from the strict adiabatic evolution, concerning the quantum states.

We have also probed the ability of this implementation to reproduce 
the adiabatic approximation
as regards to the phase.
We have performed two kinds of test.
First we have calculated the phases of the final state for two mirror-image cycles
($\dot{\alpha}(t) \rightarrow - \dot{\alpha}(t) $), still by solving 
numerically the Schr\"odinger
equation, and compared their difference $\Delta \Phi_D$   to the 
adiabatic Berry's phase, extracted
by numerical evaluation of expressions (\ref{BPphase}).
  Our result,
\be
\sin{\(( \Delta \Phi_D- (\beta^{+} (2,0; \lambda^{\star})-\beta^{-} 
(2,0; \lambda^{\star}))/2\))}= -
0.000035,   \label{NE}
\ee
  is a good confirmation that, with the chosen timing, the 
corrections to the adiabatic
approximation can be made exceedingly small. However, such an 
identity  provides a determination
of the Berry's phase only {\it modulo} $\pi$.  Using the symmetry 
$\beta^{+} (2,0; \lambda^{\star})
= -\beta^{-} (2,0;
\lambda^{\star})$, the result of our calculation actually is:
\be
\frac{1}{2}\Delta \Phi_D = \beta + \pi - 3.5\times 10^{-5}. \label{piambig}
\ee
The presence of $\pi$ should not be considered as a surprise since 
our determination of the
phase at the end of each cycle  is obtained from the argument of the 
wave function, and is therefore
defined {\it modulo} $2\pi$.

To remove the resulting ambiguity of $\pi$ on the half-difference
between mirror-image cycles, we have performed a second test.
 We compute the difference of the {\it adiabatic} phases, 
$\Delta \Phi_{adiab}$ directly from
the exact expression of the instantaneous eigenenergies of the $\vert 
2\, 0; \lambda(t)\ket$ state
during the selected cycle, by doing well-defined quadratures.   After 
comparison of both evaluations,
the result
\be
\frac{1}{2}\Delta\Phi_D = \frac{1}{2}\Delta \Phi_{adiab}  + \pi - 
2.\times 10^{-5}
\ee
merits two important remarks. First, the implementation of the
quantum cycle that we have selected, succeeds to give an 
excellent control not only of the
quantum state but also of the adiabatic phase.  Second, the $\pi $ 
increment appearing in our
numerical evaluation of Eq.(\ref{piambig}) confirms the fact that our 
theoretical evaluation of
$\Delta
\phi_D$ is defined only  {\it modulo} $2\pi$, as will also be the 
case for the experimental
determination, $\Delta \Phi_D^{exp}$. But, at the same time, we 
obtain the means to remove the {\it
modulo} $\pi$ ambiguity in the determination of $\beta$:   it is 
enough to look at the difference
$\frac{1}{2}(\Delta \Phi_D^{exp} - \Delta \Phi_{adiab})$ evaluated 
with a precision comparable to the experimental one.
  If, as we expect, this is found equal to an integer times $\pi$ with 
enough accuracy (depending on
the experimental precision, but better than 10 $\%$), then the {\it 
modulo} $\pi$ ambiguity is
suppressed from $\frac{1}{2}\Delta \Phi_D^{exp}$.

  Finally, the resulting empirical determination of the Berry's phase 
can be made free from
systematic uncertainties caused by deviations from the adiabatic 
approximation, within an accuracy
even beyond the $0.1\%$ level.  If we suppose that the time variation of $\lambda$ is realized by adjusting  the laser intensity while keeping the $\vB$ field constant $\approx$ 1~mG, the whole cycle duration amounts to $\sim$ 15 Larmor periods, {\it i.e.} 21 ms. Obviously, 
there is a compromise between
accuracy and duration: a faster $\lambda$-ramping could be employed but at the 
expense of lower accuracy.
  \subsubsection{Interferometric detection of the Berry's phase}
We can now present an explicit evaluation of the optical signal detected
for a realistic choice of the
parameters involved in the interferometric measurement. The Hamiltonian
associated with the first rf pulse is written in the laboratory frame :
\be
  H^{rf}_{lab}=\frac{w \tau _3}{2}+\frac{1}{2} \omega _1 \left(\tau
    _- e^{i w t}+\tau _+ e^{-i w t}\right).
  \ee
  In this subsection we shall use a unit system such that $\hbar =1$ and we
set $w=2\pi \Delta W$. The symbols $\tau_1, \tau_2, \tau_3$ stand for
the familiar Pauli matrices and $\tau_{\pm}= (\tau_1 \pm i\, \tau_2)/2 $.
With these notations  $ \omega _1 $, represents the coupling of the rf
frequency field $B_{rf} \hat z\sin{ w t}$ with the magnetic dipole
transition operator from the lower state $\vert F=1,m=0\ket$ to the
upper state  $\vert F=2,m=0\ket$, and the mixed states are described by a
two-component spinor $\Psi(t)=\{ X_2(t),X_1(t) \}$.  A straightforward
computation performed within the frame rotating at the angular velocity
$w$ leads at the end of the first rf pulse
($t=t_{p_1}$) to the following state vector :
\be
\tilde{ \Psi }_{rot} (t_{p_1})=
  \{ i \sin(\frac{\omega _1t_{p_1}}{2 }) , \cos(\frac{\omega _1t_{p_1}}{ 2
})\}.
\ee
  The value of $\omega_1 t_{p1}$ will be specified later on. During the
free evolution of duration T, the rf field is decoupled the rotating
frame Hamiltonian vanishes, so that $ \tilde{ \Psi }_{rot}(
t_{p1}+T)=\tilde{ \Psi }_{rot}( t_{p1})$, when the quantum cycle starts.
During the quantum cycle the
spinor component $ X_2(t)$ acquires the phase $\Phi_{2\, 0}^{\pm}$
\bea
  \Phi_{2\,0}^{\pm} = \Phi_D ^{\pm}+ \beta^{\pm}(2\, 0;\lambda^{\star}), 
\label{phi2}
   \eea
    the $\pm$ index refering to the sign of $\dot \eta$.
Simultaneously, it is also affected by the light-induced decay, but
$X_1$ is not, so we obtain
\bea
&& ~~~~~~\tilde{ \Psi }_{rot} (t_{p_1}+T+T_c)= \nonumber\\
  &&\hspace{-12mm}\{ i \sin(\frac{\omega _1t_{p_1}}{2 })\exp {\((
-\overline{\Gamma}_{dec}\, T_c /2 +i \Phi_{2\,0} \))} , \cos(\frac{\omega
_1t_{p_1}}{ 2 })\},
\eea
where $\overline{\Gamma}_{dec}$ represents the decay rate averaged over
the whole quantum cycle.

Up to now, $\omega_1t_{p_1}$ has been considered as a free parameter. 
It turns out to be
advantageous to adjust it to make the two spinor components of equal
magnitudes at the end of the quantum cycle. For this purpose, the
condition to be satisfied is $\tan{( \omega _1t_{p_1} / 2 )} = \exp {\((
\overline{\Gamma}_{dec}\, T_c /2 \))}$= 1.953 in the present example,
leading to $\arctan{\omega _1t_{p_1}/2} = 1.0975 $~rad, and a common  magnitude of 0.455, 
instead of 0.707 when there is no decay. The signal loss resulting from the 
dressed atom instability is thus a factor 2.5.  
  
  A second $ \pi/2$ Ramsey pulse,  can be applied to the
state vector and be immediately followed by the detection process. The
measured quantities are the probabilities ${\cal P}_1$ and ${\cal P}_2$
of finding the atom in states $F=1$ and $F=2$ respectively
\bea
{\cal P}_1&=&\frac{1}{2}(1-\cos{\Phi_{2\,0}}^{\pm}) \\ 
{\cal P}_2&=&\frac{1}{2}(1+\cos{\Phi_{2\,0}}^{\pm}). \label{signal}
\eea
After making the two measurements for $\dot \eta > 0$ and $\dot  \eta < 0$
and using Eq. (\ref{phi2}), one can extract $2 \beta(2 \,  0;
\lambda^{\star})$ {\it modulo} $2\pi$.
   We recall that the ambiguity of $\pi$ appearing in this 
determination of $\beta(2 \,  0;
\lambda^{\star})$ can be removed by combining this result with a 
calculation based on the knowledge of the eigenenergies.
We see that, at the price of reducing the number of detected atoms by a factor 2.5, 
 the fringe visibility can be kept very close to unity. 
 
Subsections A and B mainly refered to cold alkali atoms or trapped alkali ions, offering integer spin values  between 1 and 4. However, in the expanding 
family of laser cooled and trapped atoms there is the much less 
familiar case of chromium which, we believe deserves a special attention. 
( See section III.C). 
 \begin{figure*}
\includegraphics[height=8 cm ]{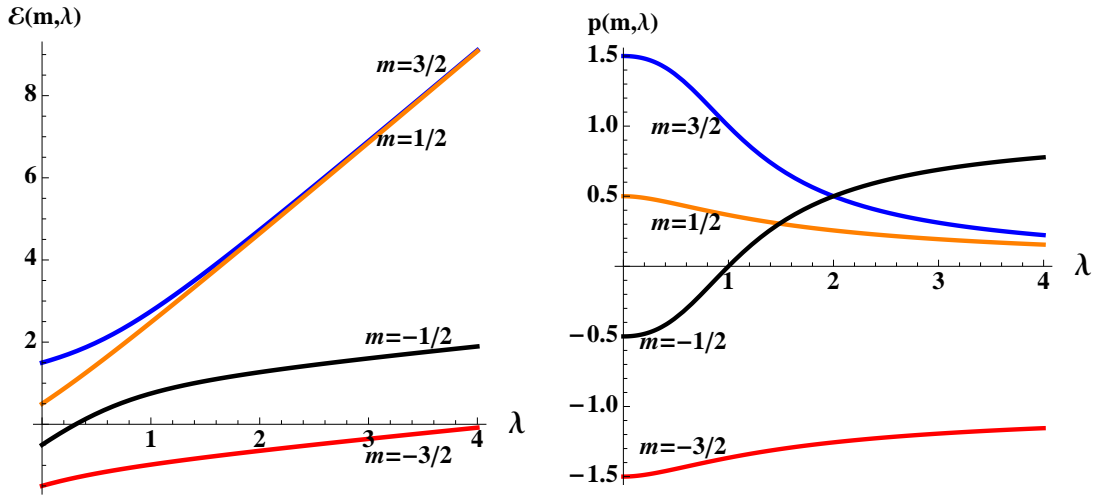}
\caption{\small (Color online) Reduced energies $\mathcal {E}(m,\lambda) $ and polarization, $ 
 p(m,\lambda)  $ for $S=3/2$ within the interval $ 0 \le \lambda \le 4 $.( The results for $\lambda <0 $  are obtained by reflexion about  the axes origin). The findings relative to
the level $ m=-1/2 $ for $2 \leq \lambda \leq 4 $ are  quite remarkable. This level remains far away from the other ones  and at the same times its  polarization $ p(-1/2,\lambda) $, giving the size of the Berry's phase, reaches values $ \geq  1$. This  indicates   a strong mixing with the level $m=3/2$. }
\end{figure*}
   \section{Berry's Phases for spins 3/2. Possible measurements}
   \subsection{ Theoretical background}
Although half-integer spins have  no fully symmetric state like  the 
$m=0 $ substate of
  integer spins, their  Berry's phases for quadratic Hamiltonians are 
interesting in their own right.
With S=3/2 we have the lowest spin state for which the dimension of the density matrix
space exceeds that of the parameter space. In order to obtain a spin 
Hamiltonian
able  to generate  the Aharonov-Anandan phase,  one would have to include
an octupole spin coupling. On the other hand, analytical  Berry's 
phase  formulas relative to the 
$ S=3/2 $ quadratic spin Hamiltonian  have already been  derived  \cite{bou2}.
In Appendix B, we give explicit expressions for $ \mathcal{E}(m,\lambda)$ and the polarization
$ p(m,\lambda)$ obtained  by rewriting  the results of ref \cite{bou2} within
the notations of \cite{bou3} and the present paper.  
The results are displayed in  Fig.5. 

The most remarkable feature apparent in this figure   
 concerns the level    $ m= -\frac{1}{2} $. 
The polarization   $ p(-1/2, \lambda) $, governing the size of 
Berry's phase, becomes   $ \geq \frac{1}{2}$ for $ \lambda  \geq 2 $ 
indicating a strong mixing with the state $ m=\frac{3}{2}$.    
At the same time,   this  level  remains at a  distance $ \simeq 1$ 
($\gamma_S B$ unit) from all other levels.
This  situation  looks favourable for an empirical determination 
of  Berry's phase   $ \beta ( \frac{3}{2}, - \frac{1}{2}) $. Moreover,  
  $ \beta ( \frac{3}{2},-  \frac{1}{2} )$  differs widely from 
 the spin one half Berry's phase $ \beta ( \frac{1}{2}, -  \frac{1}{2}) $.
This  difference provides the possibility of achieving {\it a maximum entanglement of 
  three  non-correlated spins} by the method described 
in section  VI. of reference \cite{bou3}.  More details 
about this ``holonomic"  three  Qbits entanglement are given in 
Appendix A.  
 
\subsection{Possible measurements}
There are two kinds of 3/2-spin systems for which measurements of 
Berry's phase might be
considered : i) the
$^{35}$Cl nuclei embedded inside a monocrystal matrix \cite{tycko},\cite{pines1} and ii) the
$^{201}$Hg mercury isotope in a vapor \cite{cohen1} or a cold atom optical trap  
\cite{bize} (which could apply as well to odd
isotopes of alkali-earths atoms like $^{135}$Ba and $^{137}$Ba placed in an optical dipole trap).  
Up to now in those systems, none of the experiments performed so far 
corresponds to what should be ideally
realized for testing original features of Berry's phase. 

The first experimental approach corresponds to the nuclear
quadrupole resonance (NQR) of $^{35}$Cl nuclei (I= 3/2, $\mu = 0.82\; \mu_n$) 
of an oriented axially-symmetric single crystal of sodium chlorate put inside a sample rotor 
\cite{tycko}. In this case, the
quadratic spin coupling results from
the interaction of the $^{35}$Cl nuclei quadrupole ($Q= -0.08 \times 10^{-24}$cm${^2}$) with the local axially-symmetric 
electric field gradient.  However, up to now, measurements have been performed {\it only  without} the 
linear coupling needed to lift the level-degeneracy.
To investigate the interesting range $\lambda \simeq 1$ 
 the magnetic field $\vB$ (about 10 G) should be applied along the direction $z$ of the first 
rotor axis \cite{samos},
the quadrupolar axis of the rotating monocristal being
oriented perpendicularly to $\vB$. The second rotor permits the precession of
$\vB$ about the reference axis $z$. This
set-up would be adequate for precise verifications of the Berry's
phase and its non-adiabatic corrections.
   
In the second situation relative to $^{201}$Hg, the conditions corresponding to the
hybrid  ``$\vB$-field laser-field'' situation satisfying the
$\vE\cdot\vB = 0$ hypothesis,  have been achieved in optical pumping
experiments of early 1970's \cite{cohen1}.
The quadratic spin coupling was adjusted 
as wanted over a  large range of coupling ratios $\lambda$,  
(although there was no laser available at that time) 
but, though the applied fields were time-dependent, they were far from being applied adiabatically.

   Measuring the Berry's phase  could be performed 
by preparing $^{201}$Hg atoms in a coherent superposition of two Zeeman substates,
the most interesting one being
$\Psi_{coh}= \frac{1}{\sqrt 2}\vert 3/2, -3/2\ket + \vert
3/2, -1/2\ket$ for $\lambda >0$. To this end one can start from the
pure $\vert F,m_F\ket=\vert 3/2, -3/2\ket$ substate prepared by optical
pumping and apply either a rf-frequency field or
a pulsed modulated light beam to perform transverse optical pumping
\cite{cohen1}.  The fictitious electric field can be created using {\it adiabatic application} of a 
laser beam detuned from the hyperfine lines of the $^1S_0\rightarrow\, ^3P_1$ transition.
   The resulting ground-state instability will be mild compared with 
the case of  $^{87}$Rb
   for two reasons.  First, the {\it electron}-Zeeman
coupling is replaced by the much smaller
{\it nuclear}-Zeeman coupling, so that for a given detuning, much weaker
radiation fields are needed to realize
$\lambda \approx 1$. Secondly, the life time of the $^3P_1$ Hg excited state is longer and 
the hfs splitting larger,  allowing one to select a frequency detuning of a few gigaherz, 
the ratio $\Gamma_P/\Delta {\cal W}_P$ is thus reduced by one hundred. Finally 
with all these parameters combining in a favourable way, in Hg the signal loss per cycle 
(Eq. (\ref{nodec})) becomes unsignificant. 

 After one {\it adiabatic} quantum cycle, the phase shift induced by the time-dependence of
$\vE$, expected to be $n_{\alpha}\pi(1- p(-3/2,\lambda)+p(-1/2,\lambda))$
(Eq.(\ref{BPphase})), is close to $\pi$ over the range $2\leq\lambda\leq 4$.
There are several relevant signals signals: the modulation of a transmitted resonant
probe pulse of low intensity at the
frequency $\gamma_F B \, h^{-1}({\cal
E}{(-1/2,\lambda)}-{\cal E}{(-3/2,\lambda)})$ characteristic of the dressed
atomic coherent state or the optical rotation of a linearly polarized 
beam tuned off-resonance of one hf  component of the $^1S_0-^3P_1$ 
transition. If the probe
pulse is applied at the end of the quantum cycle, the quantity to be
measured is the phase shift
of this modulation generated by the rotation of the $\vE$-field. Since 
the sign of this shift
changes with the direction of the
rotation, making two consecutive measurements, with opposite rotation
velocities, should yield the Berry's phase. When both results are 
combined, the {\it modulo} $\pi$
ambiguity has to be resolved in the same way as indicated for the 
$^{87}$Rb interferometry
experiment (cf. Sec. IV).  

A different measurement scheme could 
exploit optical pumping with
polarization modulated light, which can generate selectively 
high-order coherences \cite{budker,budker2}.

 \section{ Anandan-Aharonov  geometric phase for S=1 }
    In this section we shall deal with the 
    AA geometrical phase, still unobserved in the case of a spin-1. This phase is generated  by quantum cycles 
    along  closed circuits drawn upon the  space   $ \vE(\rho)$  of the pure state density
    matrices  $ \rho(t) =\vert\Psi(t) \ket \bra \Psi(t) \vert $. During the cycle, at any instant $t$, the
    parallel transport condition:  $ \bra\Psi(t) \vert \frac{d}{dt} \Psi(t) \ket=0 $
   has to be exactly satisfied.
    After considering the general form of the parallel transport Hamiltonian \cite{bou1}, we shall show that the hybrid B-field light-shift Hamiltonian can be tailored to suit to this form by adjusting the time-dependences of its parameters. This allows us to suggest a method for performing the measurement of the AA phase for spin-1, valid for example in the case of $^{87}$Rb atoms in the $F=1$ hf state.
   \subsection{ Aharonov-Anandan's versus Berry's phase as physical objects }
    In the spin-1 case the density matrix   is completely determined by the knowledge 
    of the polarization vector $  Tr(  \rho(t)  \vS )/\hbar $  and the ``alignement"
    tensor $A_{i,j}= Tr(\lbrace S_i, S_j \rbrace)/\hbar^2  $.  By performing
    an appropriate rotation upon the spin system 
    $ R(t)= R(\hat z , \varphi(t) ). R(\hat y , \theta(t) ).  R(\hat z , \alpha(t)) $     
     involving the three Euler angles,  the tensor $ A_{i,j} $ can be put under a
     diagonal form:
     $ \mathcal{A}= \hat z\otimes \hat z +
     \frac{1}{2} (1+\sin{\zeta}) \,\hat x\otimes \hat x +
     \frac{1}{2} (1-\sin{\zeta}) \,\hat y\otimes \hat y $
        with the polarization lying  along the z axis, $  \vp= \cos(\zeta) \, \hat z $ where 
        $ -\pi /2 < \zeta <\pi/2 $. The angle $\zeta$, together with the 
        Euler angles,  provide a system of coordinates   for $ \vE(\rho)$   which,
        by construction, is isomorphic to  $\vC P^2 $. In Reference \cite{bou1}, G. Gibbons and one of us (CB),  we have derived  the explicit expression of the  AA  geometrical phase in terms of the above set of coordinates:
        \be
        \beta_{AA}= \oint_{\mathcal{C}} \cos(\zeta )\, \cos ( \theta )\, d \varphi 
 - (1-\cos(\zeta))\, d \alpha. 
         \label{betaAA}
         \ee
   The closed loop $\mathcal{C}$ is  specified by  the following constraints within the time 
   interval $ 0 \leq t \leq  T_c$:
    $\zeta(T_c)=\zeta(0), \, -\frac{\pi}{2}  <\zeta(t)  <\frac{\pi}{2},\,  
   \theta(T_c)=\theta(0),\,  0<\theta(t)<\pi  $ and
   $ \varphi(T_c) =\varphi(0)+ 2 n_{\varphi} \pi, \,
   \alpha(T_c) =\alpha(0) + n_{\alpha} \pi$, where  $ n_{\varphi} $ and  $ n_{\alpha}$
   are non-vanishing integers.  By setting $ \lambda =  2 \tan( \zeta )$ and  
   using the method described  in our previous work 
   \cite{ bou3}, one finds easily that Berry's phase
    for $ S=m=1 $:  $  \beta(1,1)=  \beta_{AA} \;  mod (2 \pi ) $.
    
 We would like to analyse the respective behaviour of $  \beta(1,1)$ and $\beta_{AA} $
 in the limit $ \zeta= \frac{\pi}{2}+0^{-} $ or equivalently $ \lambda \rightarrow +\infty.$
 It is easily seen  that our set of coordinates  for   $\vC P^2 $ is singular for  $ \zeta= \frac{\pi}{2}$,  since  $ \vp $ is vanishing  and the alignement tensor $ \mathcal{A}$ reduces to 
 $ \hat z\otimes \hat z +\hat x\otimes \hat x $. Within such a  density matrix configuration it is not possible  to define  the Euler angle $\alpha$;  this is like the longitude for  spherical  maps which is undefined at the north pole. However,  the  problem disappears when $  \zeta= \frac{\pi}{2}-\epsilon $, $\epsilon$ being a small positive real, so that  $ \beta_{AA} $,  which is designed to be free  of non-adiabatic corrections,  is   well-defined  by the closed loop integral (\ref{betaAA}), provided the conditions listed above be satisfied.
 
  The case of Berry's phase is more delicate,  
 since   levels $ S=m=1 $  and  $ S=1 \,, m=0 $ become degenerate in the limit  $ \lambda \rightarrow +\infty $ and  are  already  very close for $   \lambda >2$. As a typical 
 example, let us consider  the  non-adiabatic correction  $ \Delta \beta $ given by equation 
 (73) of section V.C of \cite{bou3}. It  can  be easily calculated in the present case: 
$  \Delta \beta= \int _0 ^{T_c} dt   \,
     \mu^2\,  ( \cos{\theta} \,\dot \varphi +\dot \alpha )     {p}^{(2)} (1,\lambda)$
  with $ {p}^{(2)}(1, \lambda ) = \lambda + {\cal O}\left(1/\lambda^3\right)  $. 
  The parameter $\mu$,  governing  the non-adiabatic effect is given by:  
 $ \mu=- \sin \theta \, \dot \varphi /(\gamma_S \, B )$. One sees  clearly that the 
 non adiabatic  correction $  \Delta \beta$ is literally  exploding when  $ \lambda \rightarrow +\infty $.   Quite remarkably, the  situation  is  very different in  the limit   $ \lambda \rightarrow -\infty $:
   the separation between the   levels $ S=m=1 $  and  $ S=1\,, m=0 $  grows like $ - \lambda $ 
   and  $ {p}^{(2)}(1, \lambda )={\cal O}\left(1/\lambda^3\right)$, so the non- adiabatic corrections can be ignored. 
   Concerning the AA phase,  there is practically  no difference between the two limits
    $ \zeta \rightarrow \pm \pi/2$. The above results suggest  that although  $  \beta(1,1)$ and $\beta_{AA} $  are identical mathematical objects, their actual measurement  will 
 raise very different physical problems, as this will be confirmed in the following subsections.
      \subsection{ The  spin-1 ``parallel transport"  Hamiltonian  }
  In  reference \cite{bou1}, we have constructed a Hamiltonian $ H_{\parallel }(t) $ which
  performs exactly a parallel transport around the  closed circuit  $\mathcal{C}$  drawn 
  upon $\vC P^2 $. $ H_{\parallel }(t) $ must satisfy at any time t, the parallel transport
  condition:
   $
    \bra \Psi(t)\vert H_{\parallel }(t) \Psi(t)\ket  \equiv Tr( \rho(t) H_{\parallel }(t) )=0. 
   $
 In the present section, we shall limit ourselves, for the sake of simplicity, to closed circuits
 where $ \alpha(t) $ and $ \zeta (t)$ are the sole time dependent   $\vC P^2 $  coordinates. 
 To proceed, it is convenient to introduce the ``rotated " Hamiltonian: 
$\widehat{H}_{\parallel }(t)= U^{\dag}(R(t) )  H_{\parallel }(t) U(R(t))$ with
$U(R(t))=\exp (-\frac{i}{\hbar} S_z  \alpha(t)) $. In reference \cite{bou1}, using equations (49) and (A5) it was found that $\widehat{H}_{\parallel }(t)$ takes the simple form:
 \be 
 \widehat{H}_{\parallel }(t)= \frac{\dot{\zeta}(t)}{2\hbar}  \lbrace S_x, S_y \rbrace 
 - \frac{\dot{\alpha}(t)}{\hbar} \cos( \zeta(t)) S_z^2  + \dot{\alpha}(t) S_z.
 \label{hatHpar}
  \ee
In the present context we are going to consider $ \widehat{H}_{\parallel }(t)  $  just as an ``ansatz"   and  show   that it  does possess all the desired properties. 
To this end,  we shall need
the ``rotating" frame Hamiltonian, which governs  the evolution of 
$\widetilde\Psi(t)= U^{\dag}(R(t) )\Psi(t) $:
\be
\widetilde{H}_{\parallel}(t)= \frac{1}{2 \hbar} \dot{\zeta}(t) \lbrace S_x, S_y \rbrace 
 -  \frac{\dot{\alpha}(t)}{\hbar}\cos( \zeta(t)) S_z^2.   
 \label{tilHpar}
 \ee
  From  the identity: 
  $ (S_x+i S_y)^2 -(S_x-i S_y)^2 = 2 i \lbrace S_x, S_y \rbrace $, it follows that 
  $\widetilde{H}_{\parallel}(t)  $ does not mix the state $\vert 1,m= 0\ket$
  with the states $\vert 1,m=\pm 1\ket $. 
   As a consequence, if  $\widetilde\Psi(t) $
  satisfies   the initial condition  $ \widetilde\Psi(0)= \vert 1 ,0 \ket$,
   it can be written as a 
  two-component wave function $ ( \widetilde{C}(1), \widetilde{C}(-1) )$.  
  Using standard text book formulas, it is then  easily found that to derive its evolution one can replace,
  in  $\widetilde{H}_{\parallel}(t)$,  the matrices   $ \lbrace S_x, S_y \rbrace \,\hbar^{-2}  $  and $ S_z^2 \,\hbar^{-2}$,  respectively   by the Pauli matrix $ \sigma_y  $  and the
    $ 2 \times 2  $   unit matrix. Exploiting the analogy with a Ramsey pulse, one arrives 
    at the following expression for $\widetilde\Psi(t)$:
    \be 
    \widetilde\Psi(t) = ( \cos(\frac{\zeta(t)}{2} ), \sin(\frac{\zeta(t)}{2})  )
    \exp  -i \int_0^t  d\alpha\cos(\zeta).
      \label{tilPsi}
      \ee
    Using equations (\ref{hatHpar}) and (\ref{tilPsi})  one  gets:
   $ \bra\widetilde\Psi(t) \vert\widehat{H}_{\parallel}(t) \widetilde\Psi(t)\equiv 
   \bra\Psi(t) \vert H_{\parallel}(t)\Psi(t)\ket=0$. In other words,  we have 
   shown, as announced, that  the Hamiltonian
    $H_{\parallel }(t)= U(R(t) \, \widehat{H}_{\parallel}(t)\, U^{\dag}(R(t) $  performs a 
    parallel transport along  the particular closed circuit $\mathcal{C}$  considered in this section. As a   final check, let us calculate the AA phase in the laboratory frame. Writing 
    $ \Psi(T_c)= \exp ( - \frac{i}{\hbar} ( \alpha(T_c)- \alpha(0) ) S_z ) \, \widetilde\Psi(T_c)$  and
    using the phase shift $ \arg( \widetilde\Psi(T_c)/\widetilde\Psi(0) )$, deduced from 
    equation (\ref{tilPsi}),  one arrives  at the  final expression for  the AA phase:
    \be
    \beta_{AA} = \int _0^{T_c} dt \, (  \cos(\zeta)-1) \dot \alpha (t), \label{AAphase}
   \ee
     which does agree, as
   expected, with the general formula  giving  (\ref{betaAA}) in the particular case
    $\dot \varphi =0$.      
    \begin{figure*}
 \includegraphics[ width=17cm ]{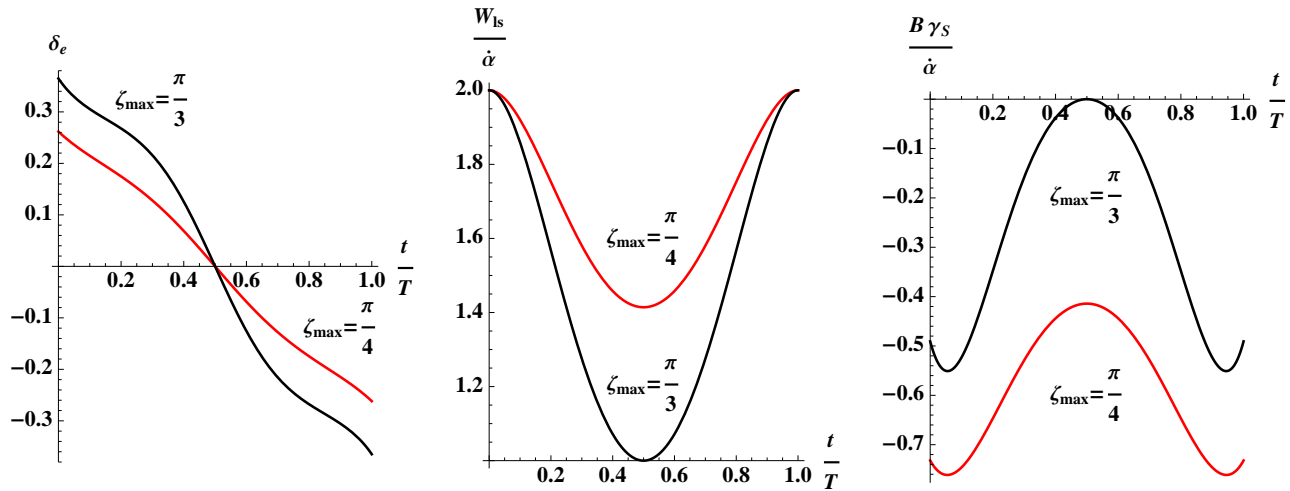}
\caption{\small (Color online) Time dependence of the physical parameters entering into the definition of the light-shift Hamiltonian which enables one to ensure parallel transport.  
 The beam helicity is $\cos{2\delta_e}$, its intensity, given by $W_{ls}$, Eq. (\ref{gammasurW}), determines the magnitude of the quadrupolar Stark coupling and $\gamma_S B+ \cos{2\delta_e}$ represents the linear dipole coupling, (for $^{87}$Rb, $F=1$ atoms the beam has to be equally detuned from the two hf excited states).   We  assume the angular velocity constant and $\zeta=\arctan{(W_{ls}/2\gamma_SB}) = 4 \, \zeta_{max}  \frac{t}{T} ( 1- \frac{t}{T} ).$  }
\end{figure*}
 \subsection{ Parallel transport  with a  light-shift Hamiltonian }
      By adding to the Zeeman Hamiltonian the light-shift Hamiltonian $\widehat{H}_{ls}$
      given by equation (\ref{Hls}), for a beam polarization 
       $ \hat \epsilon = \frac{1}{\sqrt{2}} (  \hat x +\exp ( i \chi ) \, \hat y),$
    one  gets  by a  simple calculation  a  possible candidate for an  experimental
    realization  of $ \widehat{H}_{\parallel}(t)$ :
   \bea 
    \widehat{H}_{exp}& =& \gamma_S \, B \,S_z + \nonumber \\
  &&  \hspace{-13mm}+ W_{ls}\(( \sin ( \chi ) \,S_z +
    \frac{1}{2} \cos(\chi ) \,\lbrace S_x, S_y \rbrace  - \frac{1}{2} S_z^2 \)).
      \eea
  Making the connection   with the notations of section III:
   $\chi =\pi/2 - 2 \delta_e $ and $ \vS=\hbar \vF, $ we recognize Eq. (\ref{Hhybrid-delta}) 
    where we have dropped the c-number contribution which does not contribute to the geometric phase. By  performing the identification: $\widehat{H}_{exp} \equiv 
   \widehat{H}_{\parallel }(t) $,  one obtains a set of three  relations which open 
  the road towards  experimental realization of the spin-one AA phases:
  \bea
  \sin( 2 \, \delta_e ) & =&  \frac { \dot \zeta  }{ 2 \dot \alpha  \, \cos(\zeta) }, 
  \nonumber \\
   W_{ls}& =&     2 \dot \alpha  \, \cos(\zeta), \nonumber \\
    \gamma_S \, B &=&  \dot \alpha \(( 1-  2 \cos(\zeta)  \cos( 2 \, \delta_e )\)). 
    \label{Hparex}
  \eea 
 We  have found that  physically satisfactory solutions of these equations  do exist 
 for  the following
   simple realization of the closed circuit $\mathcal{C}$, within the time interval
 $ 0 \le  t \le  T_c $, namely
 \be
  \hspace{-2mm} \zeta(t,T_c, \zeta_{max} ) = 4 \, \zeta_{max}  \frac{t}{T_c}(1 - \frac{t}{T_c} ) , \,
   \alpha(t,T_c)= n_{\alpha} \,\pi \frac{t}{T_c}, 
   \label{zetaft}
  \ee
with $  0 <  \zeta_{max} < \pi/2$  and $n_{\alpha}  $  an arbitrary integer. 

  Figure 6 represents the  three  physical quantities which determine 
 $\widehat{H}_{\parallel }(t)$, in terms of the reduced time variable  $t/T_c $, 
 for two typical values of $\zeta_{max}$, $\pi/4$ and $\pi/3$. The first curve  displays  the  angular variable 
 $\delta_e, $  which specifies   the  elliptical polarization of the light  beam.  The second
 and the third curves give in terms of  the angular   
 velocity $\dot \alpha $,   respectively the ac Stark shift  and
 the Larmor frequency  associated with the external $B$ field.  Since we have chosen the angular velocity $\dot{\alpha}$ time-independent the {\it total} magnetic field remains constant during the quantum cycle: the variation of the external field compensates the effective light-field contribution (Eq. \ref{Hparex}), therefore the total   field does not vanish during the cycle.   
 
  One should keep in mind that the ac Stark shift  introduces an instability of the 
  the  atomic ground state. For a particular  detuning configuration,  the instability of 
 an alkali ground-state with $ I=3/2$
  can be accounted for by adding to  $ \widehat{H}_{exp} $ an anti-Hermitian operator proportional  to the unit operator  $\mathbf{1} $:
  $ \widehat{H}_{exp}= \widehat{H}_{\parallel }(t)- 
 \frac{i}{2}\Gamma_{dec} \mathbf{1} $
 with $\Gamma_{dec}= 4 \; \frac{\Gamma_{P} } {2 \pi  \Delta{\cal W}_P}  \, W_{ls} $,  
  $\Gamma_{P}$ and $ \Delta{\cal W}_P$ being the decay rate and the
hyperfine splitting of the $ P_{1/2} $ state, (Eq.\ref{gammasurW}). It is clear  that such an  instability does not affect the phase shift of the surviving  atoms at the end of the quantum cycle.
 Using equation  (\ref{Hparex}) and the explicit expression of  the AA phase (\ref{AAphase}), one can 
 calculate the average value of $\Gamma_{dec}(t) $  over the AA  cycle as follows:
\be
 T_c\,  \langle\Gamma_{dec}\rangle =  \int_0^{T_c}  dt \,  \Gamma_{dec} (t)= \frac{8\,\Gamma_{P} } {2 \pi \Delta {\cal W}_P} \(( \beta_{AA} +n_{\alpha} \, \pi \)).
\label{AAloss}
\ee
        To end this section, we quote the values of the  AA  phases when 
     $\zeta_{max}= \pi/4, \; \pi/3 $ and  $  n_{\alpha} =1 $ as well as the values of $  T_c\,  \langle\Gamma_{dec}\rangle$ in the case of $^{87}$Rb,   
   \bea  \zeta_{max}&=&\frac{\pi}{4} \rightarrow \beta_{AA}=- 0.4969, \;\; \;
   T_c\,  \langle\Gamma_{dec}\rangle = 0.148,\nonumber \\
    \zeta_{max}&=&\frac{\pi}{3 }\rightarrow \beta_{AA}= -0.8567,\;\; \;
    T_c\,  \langle\Gamma_{dec}\rangle = 0.128.  \;\;\;
    \eea
 As expected,  $\beta_{AA} $ is   independent of the time scale $T$.  
   The fact that   this scaling property holds also  for $ T_c\,  \langle\Gamma_{dec}\rangle$ is less evident.There is  a strong contrast with the case of  Berry's phase measurement, where  much higher atom losses  are  unavoidable  if one wants to keep the non-adiabatic corrections below the $ 0.1 \% $ level. 
\subsection{Realization of the parallel transport Hamiltonian on $^{87}$Rb atoms}
 We suggest the realization of AA quantum cycles on spin-1 for the case of $^{87}$Rb atoms in the $F=1, m_F=1$ hf state,  when  the sole time varying parameters are  $\zeta$ and $\alpha$.  
In practice this could be achieved by a well-orchestrated time-variation of the physical parameters $W_{ls}(t), \, B(t), \, \delta_e(t)$ and $\alpha(t)$ which characterize the B-field light-field configuration with $\vB $ colinear to the beam. 
Measurements can be done with a laser beam detuned midway from both hf components starting  from the $F=1$ ground state ($\Delta_{1,1}=-\Delta_{1,2}= 2\pi \, \Delta {\cal W_P}/2$) and with the magnitudes of $B$ and $W_{ls}$ adjusted for giving to both $B$- and $E$-couplings comparable magnitudes (see Fig. 6). 
 
A precise interferometric measurement of the AA phase seems possible, the interferometric cycle being  organized in  a way very similar to the one described on Fig.3. Now, the first Ramsey pulse prepares a coherent superposition of $F=1,m_F=1$, submitted to the AA cycle, with the $F=2, m_F=0$ state which serves as a reference. However, the features of the quantum cycle itself will strongly  differ, for comparable physical conditions ({\it i.e.} average field magnitude and light intensity).
Instead of the progresssive application of the time dependent Hamiltonian required for measuring Berry's  phase precisely, one can apply the ``parallel transport'' Hamiltonian {\it suddenly, without causing any alteration of the AA phase}. The total duration of the quantum cycle can thus be reduded by one order of magnitude. 
Thus, the atom loss during the quantum cycle is also greatly reduced.  

 To best illustrate the flexibility available in the choice of physical parameters allowing one to perform AA-cycles we end by a summary of the relations  expressing  how  the cycle duration $T_c$ is linked to the laser intensity averaged over one cycle on one hand, and to the effective magnetic field on the other hand  (assuming the $\zeta$ and $\alpha$ time-dependences given by expressions (\ref{zetaft}), with $\zeta_{max} = \pi/4$): 
\bea
&&  T_c \;\; \langle W_{ls} \rangle = 2 (\beta_{AA} + \pi) = 5.28943, \\
&& \gamma_S B_{eff}\; T_c = \pi 
\eea
In addition,  we underline that, remarkably, {\it the atomic loss  per cycle, $T_c\,  \langle\Gamma_{dec}\rangle$, remains constant, whatever the absolute cycle duration} (see equation (\ref{AAloss})).      

\section{Summary and Perspectives} 
The main purpose of the present paper is to suggest methods for measuring the Berry's and AA quantum phases, in realistic experimental conditions,
  satisfying    the physical requirements  formulated in our theoretical work \cite{bou3,bou1,bou2}.  
 The spins are supposed to be non-linearly coupled to time-dependent electromagnetic fields (possibly effective ones) involving the simultaneous contributions of a linear and a quadrupole coupling. We have avoided the situations leading to degenerate eigenvalues, from which result non-Abelian Berry's phases. Important simplifications in Berry's phase calculation result from our assumption that the two effective fields involved, electric and magnetic, are orthogonal; as we have shown, the problem  becomes then endowed with several symmetry properties which make it tractable for arbitrary spin values. However, Berry's quantum cycles, performed in the Hamiltonian parameter space, require fulfilment of the adiabatic condition at any time of the spin evolution, a condition not easily satisfied in experiments on integer spin values $>1$  where the expected original features (Sec.II) deserve confrontation with our predictions.  
 On the other hand, the AA phase is  associated  directly with a quantum cycle in the density matrix space. The  so-called  ``parallel transport" condition   is satisfied directly  without  the help
of the the adiabatic approximation. The construction of 
the  spin-1 ``parallel transport"  Hamiltonian - described in a simple case in section VI.B - is a rather difficult task. Instead of a fixed number of external physical parameters it involves $4S$ parameters necessary to define  a closed circuit in the density matrix space. This construction was performed initially in ref.\cite{AAphas} for spin $  S= \frac{1}{2} $ and, later on, in  ref.\cite{bou1} for $ S=1 $, where the physical parameters involved are then the polarization vector and the alignment tensor. Detailed measurements have been performed in ref.\cite{pines} for spin $ S= \frac{1}{2}$  but no such investigation exists for $S=1.$
Our goal here has been to find how to go beyond the limitations encountered so far by experiments. To this end we describe concrete experimental situations, mainly chosen in  the field of atomic physics. We propose 
\begin{itemize}
\item{ \it }a realization of Berry's cycles for spins $>$ 1/2 having   quadrupole and dipolar couplings
 to external fieldd with  non-adiabatic correction below the $0.1 \%$ level.
 \item {\it}a  realization of  parallel transport quantum cycles  on  the spin $S=1$ density matrix 
 leading  to a  measurement of  the AA phase.
 \end{itemize}
As seen in Sec.III, the total angular momentum of atoms in their ground state are good candidates for  playing the role of isolated spins. 
 For both Berry's and AA cycles, a convenient experimental tool for coupling the spins non-linearly to external fields happens to be the ``B-field light-shift" Hamiltonian.  There are several variants: the quadrupolar coupling is realized thanks to the ac Stark shift induced  by the  linearly polarized light field, while a dipolar coupling of comparable magnitude can be ensured either by an an external magnetic field (Eq.(\ref{lambhyb})) or   the circular  polarization of the light field (Eq.(\ref{Hhybrid-delta})). 
  There is one drawback: the ac Stark shift induces an instability of the ``dressed'' ground state.
 The more severe problem lies  in the fact that 
 the quadrupole to dipole magnitude ratio $ \lambda $ acquires an imaginary part 
 which invalidates the derivation of Berry's phase given in ref. \cite{bou3}. 
In the case of alkali atoms (say $^{87}$Rb), we have found a simple remedy to get rid of  this
 unwanted imaginary  contribution: it is to tune the dressing beam frequency in such a way that
the detunings  $\Delta_{21}$ and   $\Delta_{22 }$ relative to the two transitions 
$ F=1\,; 5S_{\frac{1}{2}} \rightarrow  \mathcal{F}=1,2 \,; 5P_{\frac{1}{2}}$  satisfy the simple
relation $\Delta_{21}+\Delta_{22}=0$. Concerning the instability of our isolated spin candidate,
this puts a lower limit upon the duration of Berry's quantum cycle which cannot be too short 
if one wishes to keep the non-adiabatic corrections below the level of $  0.1 \%$. A realistic proposal to solve this delicate problem is given in subsection IV.B.   

To explore the still non-revealed Berry's phase properties expected for a spin of two, remarkable in the case $m=0$, we suggest (Sec.IV) a variant of a Ramsey interferometry experiment made on the clock transition of cold $^{87}$Rb atoms. Between the two Ramsey pulses the free evolution of a coherent superposition of the two $m=0$ hyperfine substates is interrupted by implementation of one quantum cycle in the upper state. The atoms interact with the off-resonant laser field, whose linear polarization rotates of $\pi$ around $\vB$. 
 The cycle is organized in three steps i) the laser intensity is ramped up from 0 to its maximum, with its time-derivative tailored to fit a Blackman pulse; ii) the polarization is rotated with a time-dependent angular speed described by a Blackman pulse;  fine tuning the laser intensity makes it possible to adjust the two-coupling ratio at its ``magic'' value depending on the instantaneous angular speed;  iii) the laser intensity is ramped down to zero in the time-reversed way of step i). Measurement of the phase is repeated for the ``mirror-image'' cycle (opposite rotation speed), the half-difference is expected to provide the adiabatic Berry's phase. 
 We have simulated the experiment by performing the numerical resolution of the Shr\"odinger equation in the rotating frame, allowing us to extract the value of the phase accumulated under the effect of the rotating E-field. We have found the result predicted by Berry's phase expression with a deviation of  only a few $10^{-5}$. We conclude that, within this chronology, the  empirical determination of the Berry's phase can be made free from any systematic uncertainty caused by deviations from the adiabatic approximation, within an accuracy well better than $0.1\%$.  In addition, the selected timing contributes to minimize the cycle duration, once given the magnitude of non-adiabatic correction tolerated for one measurement. It is found that the problem raised by the state instability inherent to the light-induced quadratic coupling, is expected to cause only mild experimental difficulties. 
 
As an example of Berry's phase for half-integer spins we have chosen $S=3/2$. Our predictions can be experimentally verified on $^{201}$Hg ground state atoms.  
For positive values of the quadrupole to dipole coupling ratio $\lambda \simeq 2$, the state $S=3/2, m=-1/2$  exhibits a   large Berry's phase in strong contrast with the $S=1/2, m=-1/2$ state behaviour. As seen in Sec.V and Appendix B, taking advantage of this $S$-dependence enables one to produce {\it maximum} entanglement between three initially non-correlated 1/2-Qbits in the state $m=-1/2$.

 As Sec.VI demonstrates, our goal aiming at the exploration of AA cycles by realizing parallel transport of spins $S=1$ looks  achievable.  Thanks to precisely adjusted {\it time dependences of the light-shift Hamiltonian parameters}, (intensity,  helicity of the beam, polarization rotation speed and magnitude of the magnetic field colinear to the light beam), it is possible to satisfy the parallel transport condition at any instant $t$ of the cycle. As a result the trajectory followed by the spin-1 quantum state during its cyclic evolution is a closed loop composed of geodesic segments drawn on the spin-1 state space, {\it  i.e.} the complex projective plane $ \vC P^2 $. This provides new possibilities for exploring empirically the non-trivial geometrical properties of this four-dimensional space.
 
 From a physical point of view, the main interest of the AA geometrical phase is that it could 
 be generated by using as a $S=1$ candidate the $F=1$ hf level of the ground state of $ ^{87}$Rb, which
 is the building block of one of the most popular ``optical" crystal. As for Berry's phase there is a large flexibility available on the cycle duration provided correct scaling is applied to the  other parameters, beam intensity and magnetic field. 
The advantage of the spin-1
 AA phase compared to Berry's is twofold.
\begin{itemize}
\item {a.}
 All limitations coming from the necessity of getting rid of the non adiabatic corrections
disappear.
\item {b.} The instability  problem coming from the use of effective $\vB, \, \vE$ fields  are much easier to control. This follows from the remarkable fact that  
 the  atom loss  during the cycle duration  $ T_c $ depends only on the value of the AA geometrical phase and not  upon  $ T_c $ (see equation (\ref{AAloss})). For a typical
value of  $\beta_{AA}$  the percentage  of atom loss per cycle   is about $ 15 \% $. 
\end{itemize}
 In the present paper we have  dealt, for sake of simplicity,  only 
with cycles involving $  \alpha $ and $\lambda$ as time-dependent coordinates. The  construction of a 
parallel transport Hamiltonian can be extended to an  arbitrary closed circuit of  $ \vC P^2 $, using the results of reference \cite{bou1}.
  
  \appendix
   \section{ Entanglement of three non-correlated  one-half spins  using Berry's
quantum  cycles}
Adapting  the method of  reference \cite{bou3},   we  introduce the  
three non-corellated  spin states   with 
$ M= \sum_{i=1}^{3} m_i= \frac{1}{2}$:
 \be
    \Phi ^{(1)}   =
    \vert -\frac{1}{2} \ket \otimes  \vert \frac{1}{2} \ket \otimes\vert \frac{1}{2} \ket,  
    \label{nocorelstate}
   \ee
$ \Phi ^{(2)}$ and 
$ \Phi ^{(3)}$ being obtained  by a circular permutation of the $m_i$.
These three states form an orthogonal basis  for the set of  three one half-spin states 
 with $ M=\frac{1}{2}$. The next step is to  construct three orthogonal eigenstates
 of $ \vS^2$ with $ \vS = \sum_{i=1}^{3} \vs_i $ which are linear combination of the $\Phi ^{(i)}$.
 Using the rules of adddition of quantum angular momenta, one finds that
the possible eigenvalues of   $ {\vS}^2 $,   $\hbar^2 S(S+1) $, 
 correspond to $S=\frac{3}{2}$ and  $S=\frac{1}{2}$.
 There is a unique way to construct
the state $ S=\frac{3}{2} $: one applies the operator $ S_{-}= S_x - i S_y$  upon the
state $S=M=\frac{3}{2}$, {\it i.e.} 
$\Psi_{\frac{3}{2}\;\frac{3}{2}}= \vert \frac{1}{2} \ket \otimes  \vert \frac{1}{2} \ket \otimes
    \vert \frac{1}{2} \ket   $. One obtains 
immediately:
\be
\Psi_{\frac{3}{2}\,\frac{1}{2}}=\frac{1}{\sqrt{3}} ( \Phi ^{(1)} +\Phi ^{(2)}+\Phi ^{(3)}  ).
\ee
This state is  invariant under all permuations of the 3-spin states. One has now to construct  two orthogonal  states  with $ S=M=\frac{1}{2}$, denoted $\Psi^{n}_{\frac{1}{2}\;\frac{1}{2}} $, $(n=1,2) $,
which differ by their symmetry under the  permutations of the three spins.
Ignoring  for a moment the orthogonality condition, it is  easy  to obtain two such states, linearly independent,
 $\Phi^{(1,j)} =(\Phi ^{(1)} - \Phi ^{(j)})/\sqrt{2}$
with $j=2,3$.  By  introducing in $\Phi^{(1,j)}$ the explicit expression
 of the states  $\Phi^{(j)}$, one finds, by applying the rising operator $S_{+}=S_x + i S_y$, that indeed, 
 $S_{+}\Phi^{(1,j)} = 0$.
It is then easily seen that two orthogonal states  
  $\Psi^{n}_{\frac{1}{2}\;\frac{1}{2}} $  are given, up to a
normalization factor,   by the sum and the difference  $ \Phi^{(1,2)}\pm \Phi^{(1,3)}$:
 \be 
 \Psi^{1}_{\frac{1}{2}\;\frac{1}{2}} =
\frac{1}{\sqrt{6}} (  2\Phi ^{(1)} -\Phi ^{(2)}-\Phi ^{(3)};
   \Psi^{2}_{\frac{1}{2}\;\frac{1}{2}}=  \frac{1}{\sqrt{2}}( \Phi ^{(2)}-\Phi ^{(3)}).
 \ee
 It is now  a matter of simple algebra to write the non-correlated state $ \Phi ^{(1)}$  as a linear 
 combination of the three  above angular momentum eigenstates:
 \be
\Phi ^{(1)}= \frac{1}{\sqrt{3}}(\Psi_{\frac{3}{2}\,\frac{1}{2}}+
\sqrt{2}\,  \Psi^{1}_{\frac{1}{2}\;\frac{1}{2}} )
\label{PhivsPsi}
\ee  
  An important feature of  the  three states
    $ \Psi_{\frac{3}{2}\,\frac{1}{2}},  \Psi^{1}_{\frac{1}{2}\;\frac{1}{2}}, \Psi^{2}_{\frac{1}{2}\;\frac{1}{2}}$ is their symmetry properties under  permutations of the three spins. Let us introduce the fully symmetric operator
    $ \mathcal{S}= \frac{1}{6} \sum_{i=1}^{i=6 }  p_i $   where $  p_i $ is one of the 6 possible permutations
    of the three $m_i$. It is then easily verified that: 
    $  \mathcal{S} \, \Psi_{\frac{3}{2}\;\frac{1}{2}}=\Psi_{\frac{3}{2}\;\frac{1}{2}} $  while
     $ \mathcal{S}\, \Psi^{n}_{\frac{1}{2}\;\frac{1}{2}}=0$.  On the other hand,  the permutation (23) applied upon the two  $ S=\frac{1}{2}$ states gives:
      $  (23) \,\Psi^{1}_{\frac{1}{2}\,\frac{1}{2}}= \Psi^{1}_{\frac{1}{2}\,\frac{1}{2}}$ and   $   (23) \,\Psi^{2}_{\frac{1}{2}\,\frac{1}{2}}= - \Psi^{2}_{\frac{1}{2}\,\frac{1}{2}}$.  
      
    We are going to study the adiabatic evolution of the three  non-correlated,
 1/2 spin states governed  by $  H_3(t)=H( \vB(t),\vE(t) )$, which looks formally
   like  the quadratic spin Hamiltonian
   of Eq.(\ref{themodel})  discussed extensively in the present paper,
   but with a crucial difference lying  in the fact
   that  \emph{ $\vS$  is meant to  be  the total spin operator
     $\vS = \sum_{i=1}^{3} \vs_i .$} The Hamiltonian  $ H_3(t) $ is invariant  under all the
     the permutations  of the  three spins and,  as a consequence,   all its 
     non-diagonal matrix elements taken between any pair of the  three  above states
     are vanishing.  The  above result can be extended
       to  the set of  three states $ \Psi_{\frac{3}{2}\,M} ,  \Psi^{i}_{\frac{1}{2}\,M'}$  
  with  $\vert {M} \vert =\vert {M'} \vert=\frac{1}{2} $   obtained by application 
  of the raising (lowering) operators $ S_{\pm}=S_x \pm S_y $, also  permutation invariant.
 We are  then lead to  the conclusion \cite{bou3}
    that  \emph{the three  states  $ \Psi_{\frac{3}{2}\,M} ,  \Psi^{i}_{\frac{1}{2}\,M'}$  
   behave vis \`a vis  the Hamiltonian $H_3(t)$, as if they were associated
    with  isolated spins  S}. 
    
     Our Berry's cycle would be  organized in a way similar to the one we described   
     in section IV of the present paper,
     but with one difference:  the precession Euler angle $\varphi$ 
    has to be   among the  cyclic parameter in order to have a non
    vanishing  Berry's phase for $ S= \frac{1}{2}$. Otherwise it would be impossible 
    to achieve a maximum entanglement.
  As an illustration,  we shall consider situations where during the  $ \varphi $-cycles
      $ \theta $ and $\lambda$  have predefined  fixed values obtained 
      by adiabatic  ramping processes,  analogue to those discussed   in  \cite{bou3}. 
     It is then convenient to introduce the difference   $ \Delta \beta( \lambda, \theta)$  between the  $S=\frac{3}{2}$ and $S=\frac{1}{2}$  Berry's phases: 
 $ \Delta \beta ( \lambda, \theta)= \cos \theta \, \(( p(\frac{3}{2},\frac{1}{2}; \lambda ) - 
 \frac{1}{2}\)) 2 n_{\varphi} \pi $.
 
     At the end of the Berry's cycle, the initial non-correlated state   
    $\Phi ^{(1)} $, written as a linear combination of angular states 
    behaving as isolated spin systems,  
     has evolved into  the following state: 
      \bea
  \Phi ^{(1)}_{BP}(\lambda)& =&    \exp( i\,\beta(\frac{3}{2}, \frac{1}{2}; \lambda ) 
       \frac{1}{\sqrt{3}} \Psi_{\frac{3}{2}\,\frac{1}{2}} + \nonumber \\
   &&\sqrt{\frac{2}{3}} \exp( i\,\beta(\frac{1}{2}, \frac{1}{2}; \lambda )  
     \Psi^{1}_{\frac{1}{2}\;\frac{1}{2}}.
  \eea
   We rewrite  the right-hand side of the above equation  
   in  terms of  the non-correlated states $\Phi ^{(i)}$,
  and  factor out the overall phase $\chi=\beta(\frac{3}{2}, \frac{1}{2}, \lambda ) $
  in order to exhibit the Berry's  phase difference $  \Delta \beta=\beta(\frac{3}{2}, \frac{1}{2}, \lambda ) -\beta(\frac{1}{2}, \frac{1}{2}) $. We get  the final expression for our 
  candidate  for a three one half-spin entangled state:
   \bea
  \Phi ^{(1)}_{BP}(\lambda)&=& \frac{ \exp i \chi }{3} Ê(  ( 1+ 2 \exp( -i \Delta \beta) )   \Phi(1)  
   + \nonumber   \\
  & &  ( 1 - \exp( -i   \Delta \beta  ) ) (  \Phi(2)- \Phi(3) ).
  \eea
  
  To achieve   a  maximum entanglement   one must  impose   the equality of the absolute values of the two mixing coefficients, 
  $  \vert 1+ 2 \exp( -i \Delta \beta) \vert =\vert  1- \exp( -i \Delta \beta)  \vert $.
   It is easily found that this implies  the condition $  \Delta\beta= \pm \frac{2\, \pi }{3} $.
   To  get explicit results, we have chosen the typical case $ n_{\varphi}=2 $   and 
   $\theta= \frac{2\, \pi }{3} $. Using equation (\ref{pol3/2})  of Appendix B, one gets the  following value for $\lambda_{max}=  -0.784562$.
   This negative value is welcome  because it allows one to avoid  the near-crossing of the two 
   levels $ \mathcal{E }(\frac{3}{2}, \frac{3}{2}, \lambda )$ and $ \mathcal{E }(\frac{3}{2}, \frac{1}{2}, \lambda ) $ which occurs for $ \lambda  \gtrsim 1$. For $ n_{\varphi}=2$ and
    $ \theta=\frac{ 2 \pi }{3}$ the $S=\frac{1}{2}$  Berry's phase takes the 
    simple value: $\beta(\frac{1}{2}, \frac{1}{2}) = \pi \, mod(2 \, \pi) $, since we know $\Delta \beta$, this leads to $ \beta(\frac{3}{2}, \frac{1}{2}, \lambda_{max} )= - \frac{\pi}{3}  \, mod(2 \, \pi) $ and to the final form
      \bea
  \Phi ^{(1)}_{BP}(\lambda_{max})& =& \frac{1}{\sqrt{3}}( - \exp( i  \frac{\pi}{6})\,
    \vert -\frac{1}{2} \ket \otimes  \vert \frac{1}{2} \ket \otimes\vert \frac{1}{2} \ket 
    \nonumber \\
    &&  + \exp(- i  \frac{\pi}{6}  )\,\vert \frac{1}{2} \ket \otimes  \vert -\frac{1}{2} \ket 
    \otimes\vert \frac{1}{2} \ket  \nonumber  \\   \vspace{10mm}
    &&  - \exp(- i  \frac{\pi}{6}  )  \, \vert \frac{1}{2} \ket \otimes  \vert \frac{1}{2} \ket 
   \otimes\vert -\frac{1}{2} \ket ). \label{BPstate}
    \eea
    
      The initial non-correlated    state   $\Phi ^{(1)}$ has  been transformed, 
   at the end of this specially designed Berry's cycle,
 into a correlated state with  a maximum  entanglement. 
 \section{ Berry's phase quantum cycles generated by  S= 3/2 non linear
Hamiltonians}
  We give here basic formulas  for  the  Berry phases  relative to
a spin $S =\frac{3}{2}$. They have been adapted from reference  \cite{bou2}
in order to make them  compatible with the notations of the present paper.
We begin by the reduced   Hamiltonians
${\cal H}_{even} $ and  ${\cal H}_{odd} $  connecting states with $
(-1)^{3/2-m} =\pm 1$
respectively:
\bea
{\cal H}_{even}(3/2,\lambda)&=&
\left(
\begin{array}{cc}
 \frac{3 \lambda }{4}+\frac{3}{2} & \frac{\sqrt{3} \lambda }{2} \\
 \frac{\sqrt{3} \lambda }{2} & \frac{7 \lambda }{4}-\frac{1}{2}
\end{array}
\right)
   \nonumber\\
 {\cal H}_{odd}(3/2,\lambda)&=&
\left(
\begin{array}{cc}
 \frac{7 \lambda }{4}+\frac{1}{2} & \frac{\sqrt{3} \lambda }{2} \\
 \frac{\sqrt{3} \lambda }{2} & \frac{3 \lambda }{4}-\frac{3}{2}
\end{array}
\right)
\eea
 From them one gets readily the  explicit values of $\mathcal{E}(m,\lambda)$
 and $p(m,\lambda)$ for $ m=\frac{3}{2}$ and $m=\frac{1}{2}$.
\bea
\mathcal{E}(3/2,\lambda)& =&
\frac{1}{4} \left(4 \sqrt{\lambda ^2-\lambda +1}+5 \lambda +2\right)
\nonumber \\
p(3/2,\lambda)&=&\frac{2-\lambda }{2 \sqrt{\lambda ^2-\lambda +1}}+1/2
\nonumber \\
\mathcal{E}(1/2,\lambda)& =& \frac{1}{4} \left(4 \sqrt{\lambda ^2+\lambda
+1}+5 \lambda -2\right) \nonumber \\
p(1/2,\lambda)\
&=&\frac{\lambda +2}{2 \sqrt{\lambda ^2-\lambda +1}}-1/2
\label{pol3/2}
\eea

  The formulas for $ m <0$  are easily obtained  from the reflexion law
derived in
  \cite{bou3} section III:
 $$
  \mathcal{E}(-m,\lambda)=   -\mathcal{E}(m,-\lambda) \, , \,
p(-m,\lambda)=-p(m,-\lambda).
  $$

 \end{document}